\documentclass[10pt,newlfont,pra,floats,amssymb,amsmath,showpacs,showkeys,
superscriptaddress,aps,twocolumn,letter]{revtex4}
\usepackage{color} 
\usepackage{graphicx}
\usepackage{here}
\usepackage{amsmath}
\usepackage{exscale}
\usepackage{lscape}
\usepackage[sort&compress]{natbib}
\def\ft{\hspace{0.1cm}}

\def\an1{$n^{(1)}(r)$}
\def\ea{{\it et al.}}
\def\bn2{$n^{(2)}(r)$}
\def\rb85{${}^{85}$Rb}
\def\cn01{$n_0^{(1)}(r)$}
\def\Rb87{${}^{87}$Rb}
\def\hs{\hspace{0.4cm}}
\def\dn02{$n_0^{(2)}(r)$}
\def\bls{\baselineskip -0.2cm}
\def\nac3{$na_c^3$}
\def\xna3{$na_{HS}^3$}
\def\n0{$n_0$}
\def\cn01{$n_0^{(1)}$}
\def\dn02{$n_0^{(2)}$}
\def\ahoeq{$a_{ho}\,=\,\sqrt{\hbar/m\omega_{ho}}$}
\def\li7{Li$^7$}
\begin{document}

\title{Condensate depletion in two-species Bose gases: A variational Quantum 
Monte Carlo study}
\author{A. R. Sakhel}
\affiliation{Faculty of Engineering Technology, Al-Balqa Applied University 
Amman 11134, JORDAN}
\author{J. L. DuBois}

\affiliation{University of California, Berkeley,
Department of Chemistry 19 Gilman Hall, Berkeley, CA 94720-1460}

\author{H. R. Glyde}
\affiliation{Department of Physics and Astronomy, University of Delaware, 
Newark DE 19716, USA}
\date{\today}

\begin{abstract}\footnotesize\bls
We investigate two-species Bose gases in traps with various interactions using 
variational Quantum Monte Carlo (VMC) techniques at zero temperature. The bosons
are represented by hard spheres (HS) whose diameter is equivalent to the s-wave 
scattering length in the low-energy and long-wavelength approximation. We 
explore the role of repulsive and attractive inter- or intraspecies interactions 
on the condensate properties of the mixtures, particularly the condensate 
fraction of each species as compared to the case when each species is in a 
separate trap of its own. We model the repulsive interactions by a hard core 
(HC) potential and the attractive interactions by a shallow model potential. The 
VMC density profiles and energies are evaluated at various interactions and two 
mass ratios of the species. 
\end{abstract}
\maketitle

\section{Introduction}

\hs The interest in the investigation of two-species Bose-Einstein condensates 
(2BECs) in traps has grown substantially since their first experimental 
realization in 1997 \cite{Myatt:97}. Since then, the literature on 2BECs 
in traps has exploded \cite{Esry:97,Ho:04,Shchesnovich:04,Kasamatsu:04,Ho:96,
Pu:98,Schumayer:04,Chui:03,Shi:00,Barankov:02,Cornell:98,Sinatra:00,Hall:98,
Ao:98,Trippenbach:00,Ma:06,Kevrekidis:04,Dutton:05,Mullin:06} and the interests 
are now drifting more strongly towards these mixtures \cite{Nakano:06,
Roberts:06}, which are the major theme of this paper.
 
\hs In this paper we investigate the role of inter- and intraspecies 
interactions on the properties of two-species Bose gases (2BEC) in a tight 
isotropic harmonic trap at zero temperature using variational Quantum Monte 
Carlo (VMC) methods. A tight trap enables us to simultaneously use a low number 
of particles and achieve high densities since the volume of the trapped cloud is 
much smaller than the usual size \cite{Anderson:95,Bradley:95,Sackett:98} where 
$a_{ho}\sim 10^4\AA$. Here \ahoeq\ is the trap length where $m$ is the mass of 
the boson, $\omega$ the trapping frequency, and $\hbar$ Planck's constant. We 
represent the bosons by hard spheres (HS) whose hard core (HC) diameter is 
equivalent to the s-wave scattering length in the low-energy and long wave 
length approximation. In order to describe the interactions, we use a HS 
potential for repulsive, and a shallow two-body model potential for attractive 
interactions. Basing on this, we emphasize the qualitative nature of the present
work and deemphasize comparisons with current experiments. The results stand 
alone as qualitative properties of the model system. A key point in our present  
research is that we do not use the scattering length in describing attractive 
interactions as is usually done in mean-field investigations, but rather the 
depth of a two-body model potential as justified later on. We thus vary the 
depth of the model potential and HC diameters of the bosons and investigate the 
resulting properties such as the VMC energies and density distributions. We 
particularly focus on the VMC condensate fractions and condensate density 
profiles in an -and to the best of our knowledge- unprecedented manner in the 
literature concerning mixtures of Bose gases. Another key point here is that we 
focus on the factors that enhance the condensate depletion of the 2BEC 
components. We find chiefly that the mixing of two Bose gases in a trap enhances 
the depletion of the condensates of each gas as compared to the case when either
one is in a separate trap of its own. Thus the one-component Bose gas (1BEC) in 
our paper works chiefly as a {\it reference system} to which we compare our 
mixtures. Further we find that no phase separation can occur in the case of 
attractive interspecies interactions and that two Bose gases can not be mixed 
in the case of large repulsive interspecies interactions. Some of our findings 
are similar to those of Kim and Lee \cite{Kim:02} and Shchesnovich \ea\ 
\cite{Shchesnovich:04}. Our work is particularly related to the work of Ma and 
Pang \cite{Ma:06} who did an investigation similar to ours except that they used 
repulsive interactions only, whereas we additionally use attractive interactions. 
We further evaluate the energies of the systems and check them against an 
approximate model calculation.

\hs On the theoretical side, there have been many theories and investigations. 
For example Kim and Lee \cite{Kim:02} examined the stability properties of the 
ground state of 2BECs as a function of interspecies interactions. One of the 
ground states that they found had a component localized at the center of the 
trap surrounded by the other component thereby forming a core and shell. Ho and 
Shenoy \cite{Ho:96} discussed binary mixtures of alkali condensates and found 
that the heavier of the two components always enters into the interior of the 
trap and the lighter component is usually pushed towards the edges of the trap. 
Chui \ea\ \cite{Chui:03} investigated the nonequilibrium spacial phase 
segregation process of a mixture of alkali BECs. Pu and Bigelow \cite{Pu:98} 
presented theoretical studies of a 2BEC. They showed that a mixed Bose gas 
displays novel behaviour not found in a pure condensate and that the structure 
of the density profiles is very much influenced by the interactions.

\hs On the experimental side, there also have been many investigations. For 
example Modugno \ea\ \cite{Modugno:02a} reported the realization of a mixture of
BECs of two different atomic species using potassium and rubidium by means of
sympathetic cooling. Again Modugno \ea\ \cite{Modugno:01} reported on the 
Bose-Einstein condensation of potassium atoms achieved by sympathetically 
cooling the potassium gas with evaporatively cooled rubidium. Mudrich \ea\ 
\cite{Mudrich:02} explored the thermodynamics in a mixture of two different 
ultracold Bose gases. They showed that a hot gas can be cooled to a lower 
temperature by mixing it with another colder gas. Maddaloni \ea\ 
\cite{Maddaloni:00} demonstrated an experimental method for a sensitive
and precise investigation of the interaction between two condensates. They 
studied the effects of interaction by studying two completely overlapping 
condensates and found that the center-of-mass oscillations of the two 
condensates are damped if they are interacting and otherwise if they are 
noninteracting. Matthews \ea\ \cite{Matthews:99} presented the experimental 
realization and imaging of a vortex in a two component BEC. They induced the 
vortices by a transition between two spin states by hyperfine splitting of 
$^{87}$Rb using a two photon microwave pulse. Again Matthews \ea\ 
\cite{Matthews:98} explored the dynamical response of a BEC due to a sudden 
change in the interaction strength and presented a method for the creation of 
condensate mixtures using radio frequency and microwave fields. Further they 
observed an oscillatory behaviour of the condensate sizes when the interactions 
are changed. 

\hs Although the above revealed some of the most important properties of 
ultracold mixed atomic systems in traps, an investigation of the condensate 
properties and energies is still missing. For example, what is the role of the 
hard core (HC) of the atoms in one component in determining the condensate 
fraction of the other component? In previous publications 
\cite{DuBois:01,Sakhel:02}, it has been shown that the hard core (HC) of the 
bosons plays a fundamental role in depleting the condensate of a one-species 
Bose gas in a trap. Another issue which has not been addressed before is the role 
of the mass-ratio of the bosons in a mixture in determining the condensate 
fractions and we briefly address this issue in this paper.

\hs The paper is organized as follows. In Sec. II we present the method we 
used. In Sec. III we outline our results and in Sec. IV we discuss them 
and connect to the previous literature. In Sec. V we list our conclusions 
and in Appendix A we present a model for the estimation of the energies.

\section{Method}\label{sec:method}

\hs We consider ultracold two-species Bose gases (2BEC) of $N_1$ and $N_2$ 
particles, masses $m_1$ and $m_2$, and HC diameters $a_c$ and $b_c$, 
respectively, confined in a spherically symmetric tight harmonic trap. The total
number of particles $N=N_1+N_2$ is kept fixed and we use small numbers of 
particles since larger ones increase the computational times substantially. We 
investigate the 2BECs using variational quantum Monte Carlo (VMC) methods at 
zero temperature. The program for VMC used in earlier publications 
\cite{DuBois:01,Sakhel:02} and for the one-body-density-matrix (OBDM) 
\cite{DuBois:01} has been modified to accomodate 2BECs. We shall not explain the
VMC technique as it can be found in a large number of references, rather we 
present our trial wave function and mention briefly how the particles are moved 
and how the densities are calculated.

\subsection{Hamiltonian}\label{sec:hamiltonian}

\hs The Hamiltonian of a two-component Bose gas is

\begin{eqnarray}
H\,&=&\,\sum_{i=1}^{N_1}\left(-\frac{\hbar^2}{2\,m_1}\,\nabla_{r_{1\,i}}^2\,+\,
\frac{1}{2}m_1\omega_1^2 r_{1\,i}^2\right)\,+\,\nonumber\\
&&\sum_{j=1}^{N_2}\left(-\,\frac{\hbar^2}{2\,m_2}\,
\nabla_{r_{2\,j}}^2\,+\,\frac{1}{2}m_2\omega_2^2 r_{2\,j}^2\right)\,+
\nonumber\\
&&
\sum_{i<j}V^{int}_{11}(|\mathbf{r}_{1\,i}-\mathbf{r}_{1\,j}|)\,+\,\sum_{k<\ell} 
V^{int}_{22}(|\mathbf{r}_{2\,k}-\mathbf{r}_{2\,\ell}|)\,+\,\nonumber\\
&&\sum_{m,n}\,V^{int}_{12}(|\mathbf{r}_{1\,m}-\mathbf{r}_{2\,n}|) 
\label{eq:Hamiltonian}
\end{eqnarray}

where $m_1$ and $m_2$ are the individual masses of the atoms, 
$\mathbf{r}_{\sigma 1},...,\mathbf{r}_{\sigma N_\sigma}$ are the particle 
position-vectors from the center of the trap of components $\sigma = 1$ and $2$, 
$\omega_1$ and $\omega_2$ are the trapping frequencies, 
$V^{int}_{11}$ and $V^{int}_{22}$ are the intraspecies interactions 
of species 1 and 2, respectively, and $V^{int}_{12}$ is the interspecies 
interaction.

\subsection{Units}

\hs We take length and energy in units of the trap \ahoeq\ and 
$\hbar\omega_{ho}$, respectively, where $m=m_1$ and $\omega_{ho}=\omega_1$ are 
the mass and trapping frequency of component 1, respectively. Using these 
units ($H \rightarrow H/(\hbar\omega_{ho}) = \widetilde{H}$, $r \rightarrow 
r/a_{ho} = \widetilde{r}$), the Hamiltonian (\ref{eq:Hamiltonian}) can be 
rewritten in the form 

\begin{eqnarray}
\tilde{H}\,&=&\,
\frac{1}{2}\sum_{i=1}^{N_1}\,\left(-{\widetilde{\nabla}_{r_{1\,i}}}^2\,+
\,{\widetilde{r}_{1\,i}}^2\right)\,+\nonumber\\
&&\frac{1}{2}\,
\sum_{j=1}^{N_2}\left[-\frac{m_1}{m_2}\widetilde{\nabla}_{r_{2\,j}}^2\,+\,
\frac{m_2}{m_1}\left(\frac{\omega_2}{\omega_1}\right)^2 \widetilde{r}_{2\,j}^2
\right] \,+\nonumber\\
&&\sum_{i<j}\,\widetilde{V}_{11}^{int}(|\widetilde{\mathbf{r}}_{1\,i}\,-\,
\widetilde{\mathbf{r}}_{1\,j}|)\,+
\,\sum_{k<\ell}\,\widetilde{V}_{22}^{int}(|\widetilde{\mathbf{r}}_{2\,k}\,-\,
\widetilde{\mathbf{r}}_{2\,\ell}|)\,+\,\nonumber\\
&&\sum_{m,n}\,\widetilde{V}_{12}^{int}(|\widetilde{\mathbf{r}}_{1\,m}
\,-\,\widetilde{\mathbf{r}}_{2\,n}|) \label{eq:Hamiltonian2}
\end{eqnarray}

thus introducing two ratios ($m_1/m_2$) and ($m_2\omega_2^2/m_1\omega_1^2$) into
the Hamiltonian.

\subsection{HCSW interactions}

\begin{figure}[t!]
\begin{center}
\includegraphics[width=8cm,bb=203 458 546 710,clip]{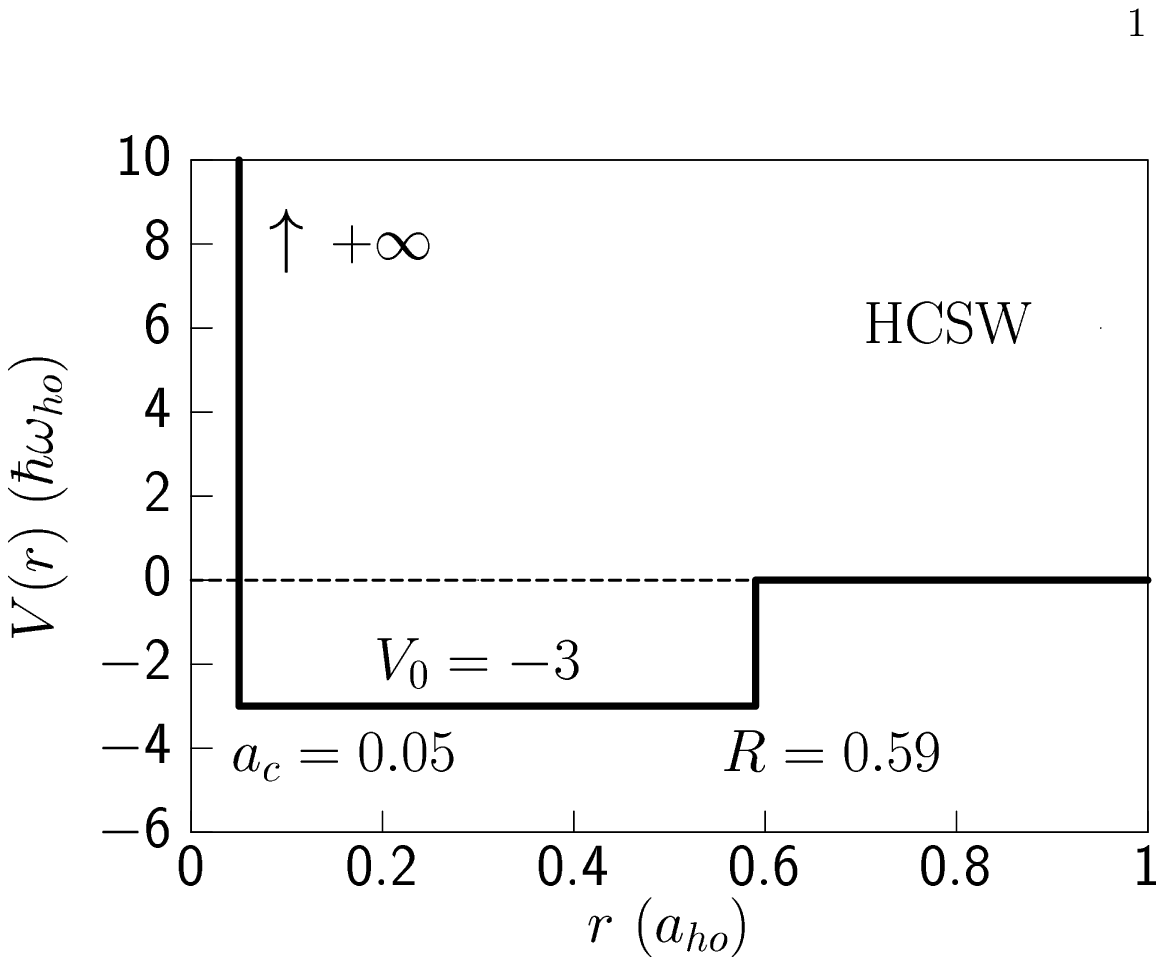}
\caption{\footnotesize\baselineskip 0.2cm HCSW interatomic potential with 
$a_c=0.05$, attractive well range $R-a_c=0.54$ and a potential depth $V_0=-3$. 
For $r\le a_c$, $V(r)$ is infinite. All lengths and energies are in trap units, 
\ahoeq\ and $\hbar\omega_{ho}$, respectively.} 
\label{fig:HCSWmodelinteraction}
\end{center}
\end{figure}

\hs We model the two-body interactions by using a hard-core square well (HCSW) 
potential. Essentially, it is a hard core plus an attractive tail added to it. 
Figure \ref{fig:HCSWmodelinteraction} shows our model potential where $V(r)$ is 
the depth and $r$ the two-body interparticle distance, all in units of the trap. 
Here for example the bosonic HC diameter is $a_c=0.05$, the depth is $V_0=-3$, 
and the range is $d\,=\,R-a_c$ which we keep fixed at 0.54. This range is of 
the same order of magnitude as that used in a previous work 
\cite{Astrakharchik:04} for another potential and we return to this point in 
Sec. \ref{sec:HCSW-properties}. In this paper we are chiefly interested in using 
the depth of the HCSW to describe the interactions and not the associated 
scattering length. Nevertheless we check the stability of the systems at the
first Feshbach resonance when $a\,\rightarrow\,\pm\infty$ in 
Sec. \ref{sec:artificial-stability} later on.

\subsection{Trial wave function} \label{sec:TrialWvfn}

\hs The general form of the trial wave function is

\begin{eqnarray}
&&\Psi_{T}(\{\mathbf{r}_1\},\{\xi_{11}\};\{\mathbf{r}_{2}\},\{\xi_{22}\};
\{\xi_{12}\})\,=\nonumber\\&&\prod_{m=1}^{N_1}
\prod_{n=1}^{N_2} f_{12}(|\mathbf{r}_{1m}-\mathbf{r}_{2 n}|) \times \nonumber \\
&&\prod_{\sigma}\left[\prod_{i=1}^{N_\sigma}g_{\sigma}(\mathbf{r}_{\sigma i}) 
\prod_{i<j}^{N_\sigma} f_{\sigma\sigma}(|\mathbf{r}_{\sigma i}-
\mathbf{r}_{\sigma j}|)\right]\,\ \label{eq:trialwvfn}
\end{eqnarray}

where $\{\mathbf{r_\sigma }\}\equiv(\mathbf{r}_{\sigma 1},...,\mathbf{r}_{\sigma 
N_\sigma})$, $g_{\sigma}(\mathbf{r}_{\sigma i})$ are single particle wave 
functions for particles of type $\sigma = 1,2$, and $f_{\sigma_1 \sigma_2}$ are
pair correlation functions for intraspecies and interspecies interactions with
variational parameters given by the sets
$\{\xi_{\sigma_1 \sigma_2}\}=\{\beta_{\sigma_1 \sigma_2}, 
\gamma_{\sigma_1 \sigma_2}, \epsilon_{\sigma_1 \sigma_2}\}$. Here the pairs 
$\{\sigma_1,\sigma_2\}$ are $\{11\}$ and $\{22\}$ for intra- and $\{12\}$ or 
$\{21\}$ for interspecies interactions. There can be several choices for the 
Jastrow functions depending on the interatomic interactions. In our case we 
constructed a flexible Jastrow function inferred from the exact solution of
two particles interacting via a HCSW potential,

\begin{widetext} 

\begin{eqnarray}
&&f_{\sigma_1 \sigma_2}(r_{ij}) = 
\left\{{\begin{array}{r@{\quad:\quad}l} 0 & r_{ij}\,\le\,a_{\sigma_1 \sigma_2} \\
\displaystyle\frac{A_{\sigma_1 \sigma_2}\,
\sin[\gamma_{\sigma_1 \sigma_2}(r_{ij}-a_{\sigma_1 \sigma_2})]}{r_{ij}} & 
a_{\sigma_1\sigma_2}\,<\,r_{ij}\,\le\,r_{\sigma_1 \sigma_2 0} \\
1\,+\,\beta_{\sigma_1 \sigma_2}^2\,
\exp[-\epsilon_{\sigma_1 \sigma_2}\,(r_{ij}-r_{\sigma_1 \sigma_2 0})^2] & 
r_{ij}\,>\,r_{\sigma_1 \sigma_2 0}\end{array}}\right. \nonumber\\ 
\label{eq:jastrowfunction}
\end{eqnarray}

\end{widetext}

where $r_{\sigma_1 \sigma_2 0}$ is the position 
of the maximum of the Jastrow function and $a_{\sigma_1 \sigma_2}$ is the HC 
diameter, where $a_{11}=a_c$, $a_{22}=b_c$, and $a_{12}=a_{21}=(a_c+b_c)/2$. The 
sinusoidal part of (\ref{eq:jastrowfunction}) is taken similar to the exact 
solution of two particles colliding inside a HCSW with relative energy 
$E_{\sigma_1\sigma_2} = \hbar^2 k_{\sigma_1\sigma_2}^2 /(2 
\mu_{\sigma_1\sigma_2})$ and HC diameter $a_{\sigma_1 \sigma_2}$ by replacing 
the HCSW wavevector $K_{\sigma_1 \sigma_2} = \sqrt{2\mu_{\sigma_1 
\sigma_2}(V_{\sigma_1 \sigma_2}+E_{\sigma_1\sigma_2})/\hbar^2}$ for each type of 
interaction of strength $V_{\sigma_1\sigma_2}$ by a variational parameter 
$\gamma_{\sigma_1 \sigma_2}$. This is in order to decouple the Jastrow functions 
from their HCSWs and to introduce some flexibility to them. Another reason for 
this replacement is that we do not know the values of $E_{\sigma_1\sigma_2}$ 
at the higher densities and we therefore allow $\gamma_{\sigma_1\sigma_2}$ to 
vary slightly in order to indirectly imply a value for $E_{\sigma_1\sigma_2}$. 
(Here $\mu_{11}=m_1$, $\mu_{22}=m_2$, and $\mu_{12} = m_1 m_2/(m_1+m_2)$.) In 
our simulations, the optimized $\gamma_{\sigma_1\sigma_2}$ is always very close 
to $K_{\sigma_1\sigma_2}$ and $\gamma_{\sigma_1\sigma_2}\ge\ 
K_{\sigma_1\sigma_2}$. Thus attractive interaction between the particles is 
included in the Jastrow via $\gamma_{\sigma_1\sigma_2}$. We then join the 
sinusoidal part at $r_{ij}=r_{\sigma_1\sigma_2 0}$ to another function which 
decays to 1 in the long range. Note then that $r_{\sigma_1 \sigma_2 0}$ is not 
necessarily equal to $R_{\sigma_1 \sigma_2}$, the edge of the HCSW for each 
interaction type, and depending on the well depth it can be either inside or 
outside the HCSW. The reason for this construction is to provide smooth Jastrow 
functions whose maxima are at interparticle distances large enough to bring the 
bosons close together. Further it is important to note that the attraction 
between the bosons is mainly caused by the Jastrow function 
(\ref{eq:jastrowfunction}), particularly by the ``bump" of the Jastrow which
is higher than 1 at $r=r_{\sigma_1\sigma_2 0}$.
The part of the Jastrow function in the range 
$a_{\sigma_1\sigma_2} < r_{ij} \le r_{\sigma_1\sigma_2 0}$ is then repulsive, 
whereas in the range $r_{ij} \ge r_{\sigma_1 \sigma_2 0}$ attractive. Note that 
in the case of only repulsive interactions the HCSW depth $V_{\sigma_1 
\sigma_2}$ and $\beta_{\sigma_1 \sigma_2}$ are set to zero. Therefore when 
$\gamma_{\sigma_1\sigma_2}\rightarrow 0$, 

\begin{equation}
\lim_{\gamma_{\sigma_1 \sigma_2}\to 0} 
\frac{\sin[\gamma_{\sigma_1 \sigma_2}(r_{ij}-a_{\sigma_1 \sigma_2})]}
{\gamma_{\sigma_1 \sigma_2}\, r_{ij}} = 
\left(1 - \frac{a_{\sigma_1 \sigma_2}}{r_{ij}} \right) 
\end{equation}

brings us back to the HS Jastrow function. $A_{\sigma_1\sigma_2}$ (and 
$r_{\sigma_1 \sigma_2 0}$) are parameters that join the Jastrow in the 
interparticle-separation range $a_{\sigma_1\sigma_2} < r_{ij} \le r_{\sigma_1 
\sigma_2 0}$ to that in the range $r_{ij} > r_{\sigma_1 \sigma_2 0}$ for the 
same slope and amplitude. For the single particle wavefunctions, we use 
Gaussians of the form

\begin{equation}
g_{\sigma}(\mathbf{r}_{\sigma i})\,=\,\exp(-\alpha_{\sigma} r_{\sigma i}^2) 
\end{equation}

where $\alpha_{\sigma}$ are variational parameters. Later on in this paper we 
shall see that even with Gaussians centered at the origin, the variational 
wavefunction (\ref{eq:trialwvfn}) is still able to describe phase separation. 
This indicates that the trial wavefunction is dominated by the pair correlation 
functions $f_{\sigma_1 \sigma_2}$ rather than the single particle functions. In 
fact the Gaussians chiefly cause the density to vanish at the edges of the cloud 
thus indirectly confining the cloud within a certain volume.

\subsection{Moving the particles}

\hs The particles are moved according to

\begin{equation}
\mathbf{r}_{\sigma i}^\prime = \mathbf{r}_{\sigma i} + 
\Delta\mathbf{r}_\sigma\cdot(\eta-0.5)
\end{equation}

where $\mathbf{r}_{\sigma i}^\prime$ are new positions, $\eta$ is a random 
number between 0 and 1 and $\Delta\mathbf{r}_\sigma$ are step size vectors which 
are adjusted to obtain optimal diffusion through configuration space-- i.e. to 
obtain a VMC acceptance rate of $\approx 50\%$. After each update of the 
particle coordinates the proposed move is either accepted or rejected 
according to the (MRT)$^2$ \cite{Metropolis:53} algorithm where the square of 
the trial wave-function is used as the probability distribution from which 
particle configurations are sampled. 

\subsection{Minimization of the variance of the energy}\label{sec:min_variance}

\hs In order to optimize the trial wavefunction, we numerically minimize the 
variance of the energy $\sigma_E$ with respect to the variational parameters of 
the trial wavefunction. The variance is given by $\sigma_E^2\,=\,\langle E^2
\rangle - \langle E \rangle^2$ where in general for any operator $O$

\begin{equation}
\langle O \rangle\,=\,
\frac{\displaystyle\int d\mathbf{r}_1 d\mathbf{r}_2 \,|\Psi_0|^2 
\,\left(\frac{O\,\Psi_1}{\Psi_1}\right)\frac{|\Psi_1|^2}{|\Psi_0|^2}}
{\displaystyle\int d\mathbf{r}_1 d\mathbf{r}_2 
\,|\Psi_0|^2\,\frac{|\Psi_1|^2}{|\Psi_0|^2}}
\end{equation}

with 
$\Psi_0=\Psi(\{\mathbf{r}_1\},\{\xi_{11}^0\};\{\mathbf{r}_2\},\{\xi_{22}^0\};
\{\xi_{12}^0\})$ and $\Psi_1 = 
\Psi(\{\mathbf{r}_1\},\{\xi_{11}\};\{\mathbf{r}_2\},\{\xi_{22}\};\{\xi_{12}\})$ 
and $\int d\mathbf{r}_1\,\equiv\,\prod_{i=1}^{N_1} \int d^3 r_{1\,i}$, 
$\int d\mathbf{r}_2\,\equiv\,\prod_{j=1}^{N_2} \int d^3 r_{2\,j}$. Here 
$|\Psi_1|^2/|\Psi_0|^2$ are weights used for the reweighting process of the 
variable $O$. $\{\xi_{\sigma_1\sigma_2}^0\}$ are the initial and 
$\{\xi_{\sigma_1\sigma_2}\}$ the optimized sets of variational parameters. In 
the Gaussians we also use the initial $\alpha_\sigma^0$ and the optimized 
$\alpha_\sigma$.

\subsection{Condensate fraction}

\hs In the systems considered here we have two condensate fractions \cn01\ and 
\dn02\ for components 1 and 2, respectively. The overall condensate fraction of 
the mixture is $(n_0^{(1)}N_1 + n_0^{(2)}N_2)/(N_1+N_2)$ but we only focus on 
the individual \cn01\ and \dn02. The condensate fraction of each component is 
evaluated by calculating the eigenvalues of the natural orbitals using the 
one-body-density-matrix (OBDM) of each component in a manner similar to a 
calculation by DuBois and Glyde \cite{DuBois:01}. By using the trial wave 
function of Sec. \ref{sec:TrialWvfn}, we evaluate the OBDM for components 1 and 
2, respectively, as follows. In order to make the equations more compact, we 
define 

\begin{eqnarray*}
&&Q_1(r_{11},\cdots) = \nonumber \\
&&\psi(r_{11},r_{12},\cdots,r_{1N_1},\{\xi_{11}\};\left\{\mathbf{r}_2\right\},
\{\xi_{22}\};\{\xi_{12}\}),
\end{eqnarray*}

\begin{eqnarray*}
&&Q_1(r_{11}^\prime,\cdots) = \nonumber\\
&&\psi(r_{11}^\prime,r_{12},\cdots,r_{1N_1},\{\xi_{11}\};
\left\{\mathbf{r}_2\right\},\{\xi_{22}\};\{\xi_{12}\}),
\end{eqnarray*}

\begin{eqnarray*}
&&Q_2(r_{21},\cdots) = \nonumber \\
&&\psi(\left\{\mathbf{r}_1\right\},\{\xi_{11}\};r_{21},r_{22},\cdots,r_{2 N_2},
\{\xi_{22}\};\{\xi_{12}\}),
\end{eqnarray*}

\begin{eqnarray*}
&&Q_2(r_{21}^\prime,\cdots) = \nonumber \\
&&\psi(\left\{\mathbf{r}_1\right\},\{\xi_{11}\};r_{21}^\prime,r_{22},\cdots,
r_{2 N_2},\{\xi_{22}\};\{\xi_{12}\}). 
\end{eqnarray*}

Hence the OBDMs are written

\begin{eqnarray}
&&\rho_a(r_{11},r_{11}^\prime)\,=\,\nonumber \\ 
&&\frac{\prod_{i=2}^{N_1}\int d^3 r_{1\,i} \prod_{j=1}^{N_2} \int d^3 r_{2\,j} 
Q_1(r_{11},\cdots) Q_1(r_{11}^\prime,\cdots)} 
{<\psi\,|\,\psi>} \nonumber \\
\label{eq:n(r)}
\end{eqnarray}

leaving out the integration over $r_{11}$ and $r^\prime_{11}$ and

\begin{eqnarray}
&&\rho_b(r_{21},r_{21}^\prime)\,=\,\nonumber \\ 
&&\frac{\prod_{i=1}^{N_1}\int d^3 r_{1\,i} \prod_{j=2}^{N_2} \int d^3 r_{2\,j} 
Q_2(r_{21},\cdots) Q_2(r_{21}^\prime,\cdots)}
{< \psi\,|\,\psi >} \nonumber \\                           
\label{eq:n(R)}
\end{eqnarray}

leaving out $r_{21}$ and $r^\prime_{21}$. Here 

\begin{eqnarray*}
<\psi|\psi> = \int d r_1 \int d r_2 
|\psi(\left\{\mathbf{r}_1\right\},\{\xi_1\};\left\{\mathbf{r}_2\right\},\{\xi_2
\};\{\xi_{12}\})|^2, 
\end{eqnarray*}

is the normalization factor and $\xi_{\sigma_1 \sigma_2}$ are the optimized 
variational parameters. Hence we extract the OBDM for each component from the 
two-body density matrix (TBDM) of the mixture by integrating out the 
contribution from the other component. In a manner similar to Ma and Pang 
\cite{Ma:06} then, each component is essentially treated as a subspecies with 
its own properties but still it is not independent of the other species as a 
result of the interspecies interactions. The interspecies interactions are 
included in the OBDM through the interspecies Jastrow function $f_{12}$. From 
the trial wave function we can verify that Eqs.\ft(\ref{eq:n(r)}) and 
(\ref{eq:n(R)}) reduce to the one-component case if the interspecies 
interactions are turned off. That is the interspecies Jastrow function $f_{12}$
becomes equal to 1 and the two components become independent of each other as 
they are now noninteracting. 

\subsection{Density profiles}

\hs The densities are calculated during a VMC run by dividing the space along 
the radial direction into spherical shells (bins) concentrated at the center of 
the trap and collecting the particles of each species in them as was done before 
by DuBois and Glyde \cite{DuBois:01}.

\section{Results}

\hs In what follows we present the results of our Monte Carlo simulations. We 
display and discuss the resulting VMC density profiles and the condensate 
fractions of our mixtures with various interactions. We further compare the 
condensate fractions of the mixtures with the condensate fractions of their 
components when either one is in {\it a separate trap of its own}. We compare 
our VMC energies with the results from an approximate mean-field model derived 
in Appendix A. We further reveal the role of the mass ratio $m_1/m_2$ in 
determining some properties of the Bose gases. The trapping frequency is set to 
be the same for both components ($\omega_1=\omega_2$) and the mass ratio 
is arbitrarily chosen to be $m_1/m_2=1.200$.

\subsection{Stability of the mixtures as compared to one-species Bose gases}

\hs During the numerical optimization of the variational parameters as explained 
in Sec. \ref{sec:min_variance}, we plot the energy $E_{VMC}/N$ versus the set of 
variational parameters $\{\xi_{\sigma_1 \sigma_2}\}$ used in our wave function.
The numerical optimization procees changes the variational parameters over 
several iterations and searches for a minimum in the energy variance which 
also leads to a minimum in the energy. After a number of iterations, we obtain 
plots such as those shown in the following figures. For example 
Fig. \ref{fig:plot.Evsalpha} displays the VMC energy $E_{VMC}/N$ vs one of the
variational parameters $\alpha_1$ for a mixture with $a_c=0.1$, $b_c=0.2$ and 
repulsive HC interactions only (upper frame) and for a mixture with $a_c=0.2$, 
$b_c=0.3$, attractive (HCSW, $V_{12}=-10.0$) inter- and repulsive HC intraspecies 
interactions (lower frame). The figure depicts clearly the presence of energy 
minima at $\alpha_1\sim 0.25$ and $\sim 2.9$, respectively. The behavior of the 
energy versus the other variational parameters is the same as in 
Fig.\ft\ref{fig:plot.Evsalpha} and all of them display energy minima. After the 
completion of the optimization process and in the final evaluation of the 
wavefunction for each system, we choose the variational parameters that correspond 
to the energy minimum, i.e., the ground state. All of our repulsive or attractive 
2BECs display energy minima as above and we can therefore state safely that our 
mixtures are stable systems. 

\hs In comparison, 
Fig.\ft\ref{fig:plot.Evs.alpha.1BEC.HCSW.N20.two.minima.special} shows the VMC 
energy against $\alpha$ for a HCSW 1BEC of 20 particles, HCSW depth $V=-6$ and 
$a_c=0.2$ using the same trial wavefunction (\ref{eq:trialwvfn}) but set for one 
component only. The figure shows a peculiar result, namely the presence of two 
equal energy minima at $\alpha\sim 1.15$ and 1.2, i.e. a degeneracy. One of the 
minima is due to the single-particle, the other due to the Jastrow part of the 
trial wavefunction. The single-particle wavefunction is connected to the 
external trapping potential and the Jastrow function to the interparticle 
interactions and generates as such the energy minima due to these potentials.
This plot has been generated from two VMC runs using different minimization 
directions \cite{Press:99} in order to ensure the presence of the two minima. We 
have seen this phenomenon in all the VMC runs for this particular system at 
various other HCSW depths. We do not understand at the present why this double 
minimum does not occur in 2BECs.

\begin{figure}[t!]
\hspace{-0.2cm}\includegraphics[width=8.7cm,bb=174 362 438 708,clip]{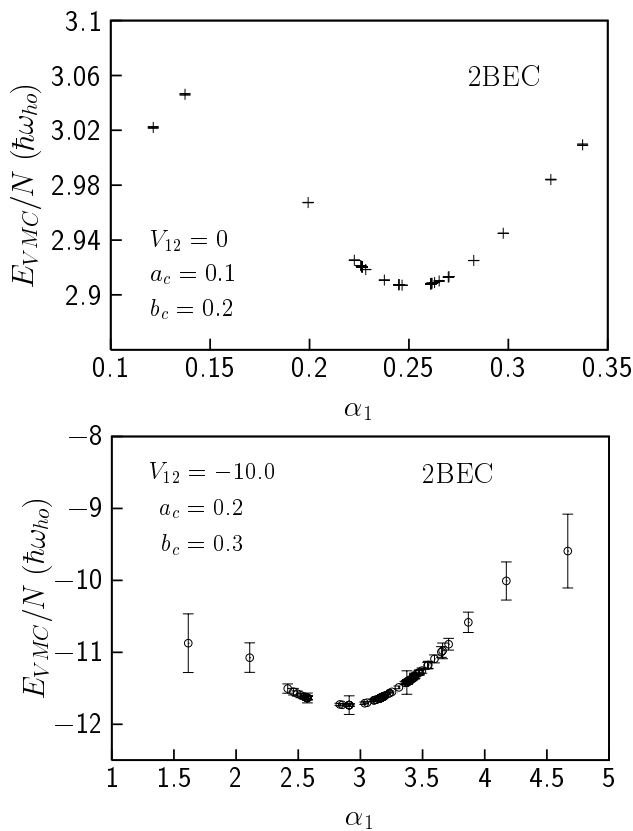}
\caption{\footnotesize\baselineskip 0.2cm VMC energy per 
particle versus the variational parameter $\alpha_1$ of the trial wave function 
(\ref{eq:trialwvfn}) for a trapped Bose gas mixture of $N_1=20$ and 
$N_2=10$ particles. Upper frame: HC 2BEC with $a_c=0.1$ and $b_c=0.2$, lower 
frame: HCSW 2BEC with $a_c=0.2$, $b_c=0.3$ and $V_{12}=-10.0$. $V_{12}$ is the 
interspecies HCSW depth.} \label{fig:plot.Evsalpha}
\end{figure}

\begin{figure}[t!]
\hspace{-0.3cm}\includegraphics[width=8.9cm,bb=167 465 545 703,clip]{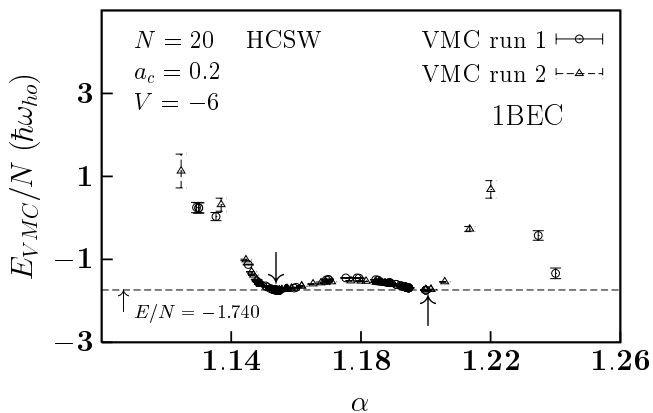}
\caption{\footnotesize\baselineskip 0.2cm As in 
Fig.\ft\ref{fig:plot.Evsalpha} but for a HCSW 1BEC with $N=20$ particles. The 
depth of the HCSW is $V=-6$ and the arrows indicate the locations of two equal 
energy minima ($E/N=-1.740$). This plot is a result of two VMC runs of the same 
system.} \label{fig:plot.Evs.alpha.1BEC.HCSW.N20.two.minima.special}
\end{figure}

\subsection{Definitions of the densities}

\hs In order to describe the density of the systems, we used $na_c^3$ with $a_c$ 
the HS diameter of the bosons for component 1 and $nb_c^3$ with $b_c$ the HS 
diameter for the bosons of component 2. We thus describe the systems by the HC 
density only, even in the presence of attractive interactions. We define the 
total VMC spacial density distributions (condensate + normal parts) by $n_1(r)$ 
and $n_2(r)$ for components 1 and 2, respectively, and in units of $a_{ho}^3$ 
where $r$ is the distance of a boson of either species from the center of the 
trap ($r=0$). Correspondingly $n_{0,1}(r)$ and $n_{0,2}(r)$ are the VMC 
condensate density distributions. The total VMC density of a 1BEC is written 
$n(r)$. In the interpretation of our results, we sometimes need to display the 
properties of both components as a function of their HC densities in a single 
plot. For this particular purpose, we use a unified term, namely \xna3\ with 
$a_{HS}$ the HS diameter $a_c$ or $b_c$, to describe the $HC$ density of either 
component at the center of the trap and \xna3\ is used then under the following 
conditions. In a single HC or HCSW 1BEC $na_{HS}^3=n(0)a_c^3$ where $n(0)$ is 
the number density at the center of the trap and $a_c$ is the HC diameter of the 
single-species bosons. Since we are dealing with more than one species, 
\xna3\ of each component has to be defined for various cases of 
interactions in the 2BECs. For systems with attractive interspecies interactions 
where there is full mixing (Sec. \ref{sec:attr.inter.rep.intra}) 
$na_{HS}^3=n_1(0)a_c^3$ for component 1 and $na_{HS}^3=n_2(0)b_c^3$ for 
component 2. In the case of repulsive interspecies interactions \xna3\ is 
the HC density of the core only since the two species 
phase-separate (Sec. \ref{sec:attr.intra.and.rep.inter} and
\ref{sec:rep.inter.and.inter.}) and it is difficult to 
define \xna3\ for a shell. In all of our interpretations, we do not consider 
\xna3\ to be the overall total density $(n_T(r)=n_1(r)+n_2(r))$ of the 
mixture at any time.

\begin{figure*}[t!]
\includegraphics[width=15cm,bb = 80 279 537 711,clip]{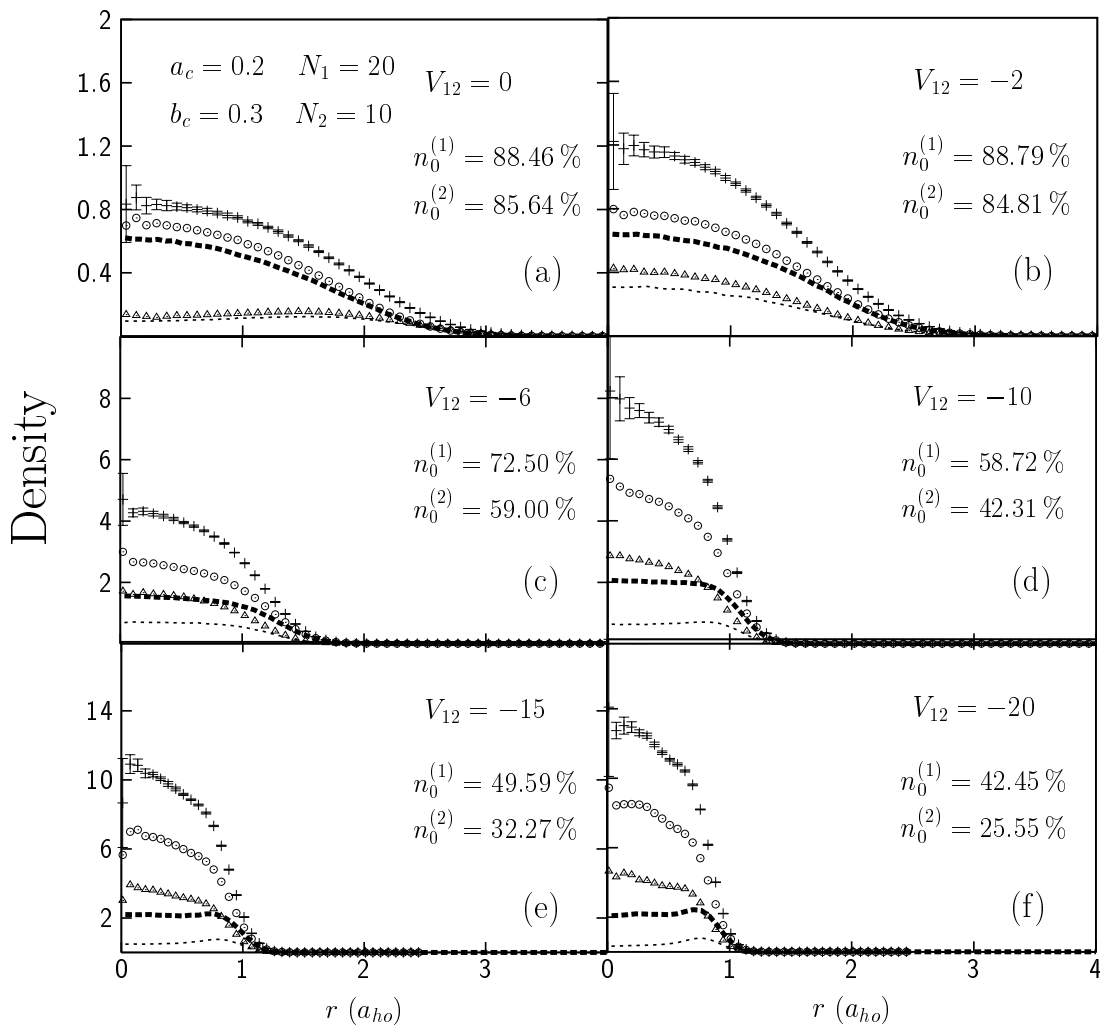}
\caption{\footnotesize\baselineskip 0.2cm VMC density profiles and condensate 
properties of 2BECs with HCSW inter- and HC intraspecies interactions. $V_{12}$ 
is the interspecies HCSW depth. Points with error bars: $n_T(r)$, open circles 
and triangles: $n_1(r)$ and $n_2(r)$, respectively, thick and thin dashed lines: 
corresponding $n_{0,1}(r)$ and $n_{0,2}(r)$, respectively.} 
\label{fig:plot.density.figures.complete.mix.stackI}
\end{figure*}

\begin{figure}[t!]
\hspace{-0.2cm}\includegraphics[width=8.8cm,bb = 92 414 495 709,clip]{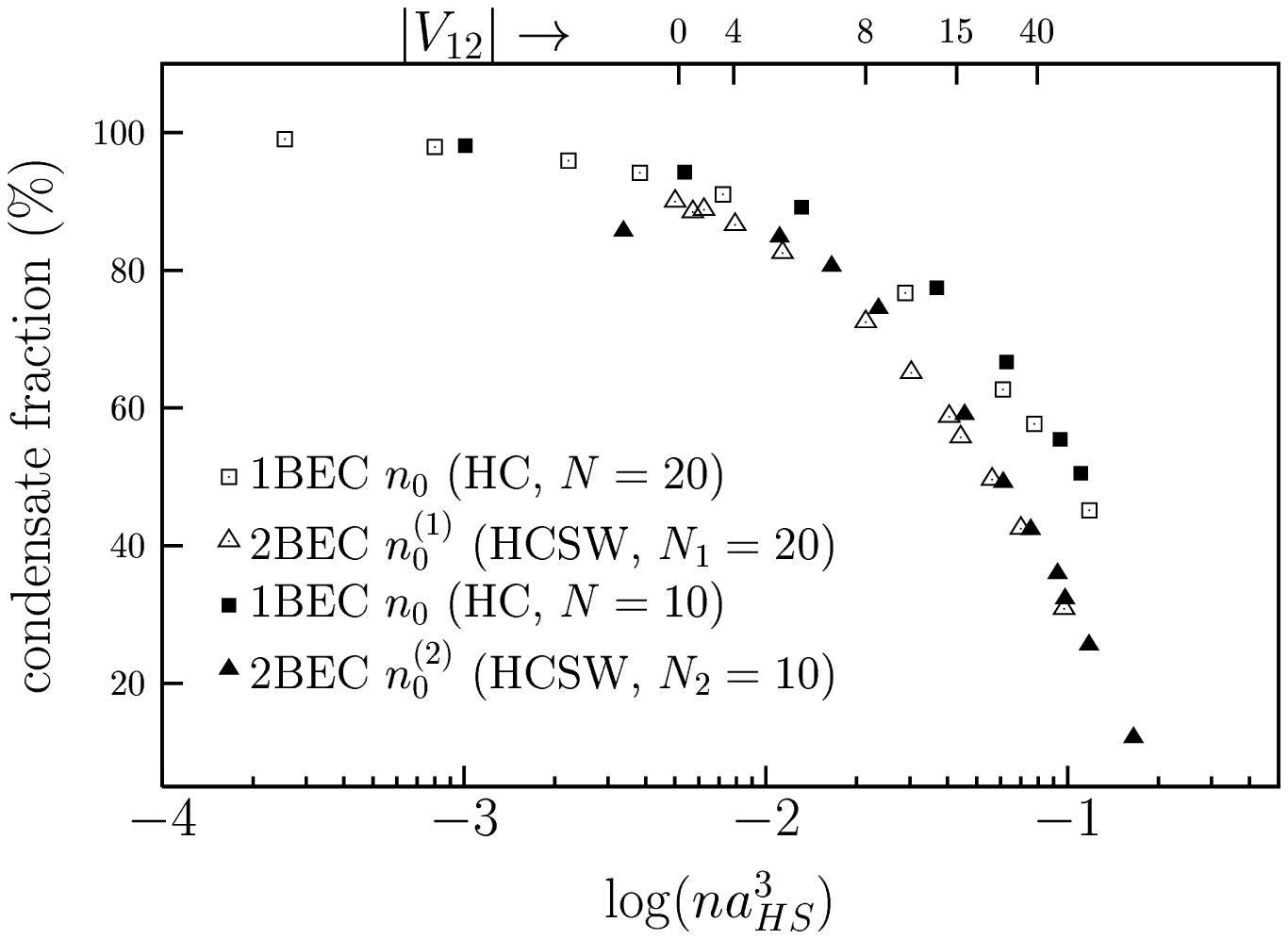}
\caption{\footnotesize\baselineskip 0.2cm Condensate fractions of HC 1BECs and 
the HCSW 2BEC of Fig.\ft\ref{fig:plot.density.figures.complete.mix.stackI} as a 
function of \xna3. Open and solid squares: ({\it reference}) HC 1BECs with 
$N=20$ and $N=10$ particles, respectively. Open and solid triangles: components 
1 and 2 of the HCSW 2BEC and $V_{12}$ is the depth of the HCSW for some of the 
points. The points are larger than the error bars.} 
\label{fig:plotn01andn0HCvsnac3}
\end{figure}

\subsection{Attractive inter- and repulsive 
intraspecies interactions}\label{sec:attr.inter.rep.intra}

\subsubsection{Density profiles}

\hs The goal of this and the following sections is to display the spacial VMC 
density profiles of 2BECs with various interaction parameters which have largely 
not been displayed before in the literature. Figure 
\ref{fig:plot.density.figures.complete.mix.stackI} displays the MC density 
profiles of 2BECs with attractive (HCSW) interspecies and repulsive (HC) 
intraspecies interactions. The points with error bars represent $n_T(r)$ of the 
mixture, the open circles represent the density $n_1(r)$ and the open triangles 
$n_2(r)$. The thick and thin dashed lines are $n_{0,1}(r)$ and $n_{0,2}(r)$,
respectively. The strength of the interspecies interactions is indicated by the 
depth of the HCSW, $|V_{12}|$, and the range of the HCSW is kept fixed at 0.54. 
In all our mixtures, here and thereon, the components have $N_1=20$ and $N_2=10$ 
particles and in this section $a_c=0.2$ and $b_c=0.3$. There is a particular 
reason for the choice of the latter large $a_c$ and $b_c$ above, this is in order 
to enable 
substantial depletion of the condensates. We keep $a_c$ and $b_c$ fixed and 
increase $V_{12}$ from 0 to $-40$ in the ``negative" sense. 

\hs The key features of 
Fig.\ft\ref{fig:plot.density.figures.complete.mix.stackI} are as follows. The 
attractive forces enable full mixing of the two components. In frame (a) 
component 2 is slightly pushed out towards the edges of the trap due to the 
repulsive interspecies interactions arising by setting $V_{12}\,=\,0$. That is 
when the attractive part of the HCSW is switched off, the HCSW changes to a 
repulsive HC potential. Then we note that although $V_{12}\,=\,0$, full mixing 
of the two components is still possible. At the instant the HCSW is ``switched 
on" as in frame (b), component 2 is {\it pulled back} towards the center of the 
trap with no remnant expulsion at the edges of the trap. The densities in frame 
(b) jump now above those in frame (a) and continue to rise as $V_{12}$ is 
increased. In frames (a) and (b) the condensate densities $n_{0,1}(r)$ and 
$n_{0,2}(r)$ are similar in shape to their corresponding total densities 
$n_1(r)$ and $n_2(r)$, but in the rest of the frames (c-f) they are not. Rather 
they obtain a flat shape in frames (c) and (d) after which they are slightly 
pushed out towards the edges of the trap in frames (e) and (f). In any case, the 
attractive forces prevent the condensate from total expulsion towards the edges 
of the trap. We anticipate that as the density rises further with $V_{12}$, the 
condensates will be pushed out further towards the surface of the cloud because 
the condensate seeks the lower density regimes of the cloud. The reason is 
because the lower cloud density at the edges of the trap causes a lesser local 
condensate depletion than the higher density towards the center. Note also that 
the total densities in frames (a-c) have a Gaussian shape, but then they divert 
from it somewhat. The densities rise also significantly with the increase of 
attractive interspecies interactions: $n_T(r)$ rises by a factor of $\approx 18$ 
from frame (a) to frame (f) as the cloud radius shrinks in size by a factor of 
$\approx 3$. 

\hs We are also able to use large attractive interspecies interactions 
($V_{12}\,=\,-40$) at the energy scale of ultracold Bose gases and still obtain 
energetically stable systems.

\subsubsection{Condensate fractions}

\hs The goal of this section is to display the effect of complete mixing on the 
depletion of each condensate in a 2BEC mixture as compared to the case when 
either condensate is in {\it a separate trap of its own} in which case it forms 
a 1BEC. For this purpose we consider the mixtures in 
Fig.\ft\ref{fig:plot.density.figures.complete.mix.stackI}. Figure 
\ref{fig:plotn01andn0HCvsnac3} displays their condensate fractions as a function 
of \xna3. The open and solid squares represent the condensate fractions \n0\ 
for HC 1BECs with $N=20$ and $N=10$ particles, respectively, which act as our 
{\it references}. The open and solid triangles display the condensate fractions 
of the mixtures \cn01\ and \dn02. The figure depicts clearly that the depletion 
of the condensates in the 2BECs is larger than the 1BECs. This reveals that 
mixing enhances the depletions of the constituent condensates due to their 
interspecies interactions. A significant feature is that \cn01\ and \dn02\ as a
function of \xna3\ coincide at the larger \xna3. We may attribute this to the 
fact that since the two components are completely mixed the system behaves 
similarly to a 1BEC.

\begin{figure*}[t!]
\includegraphics[width=15cm,bb=80 373 572 709,clip]{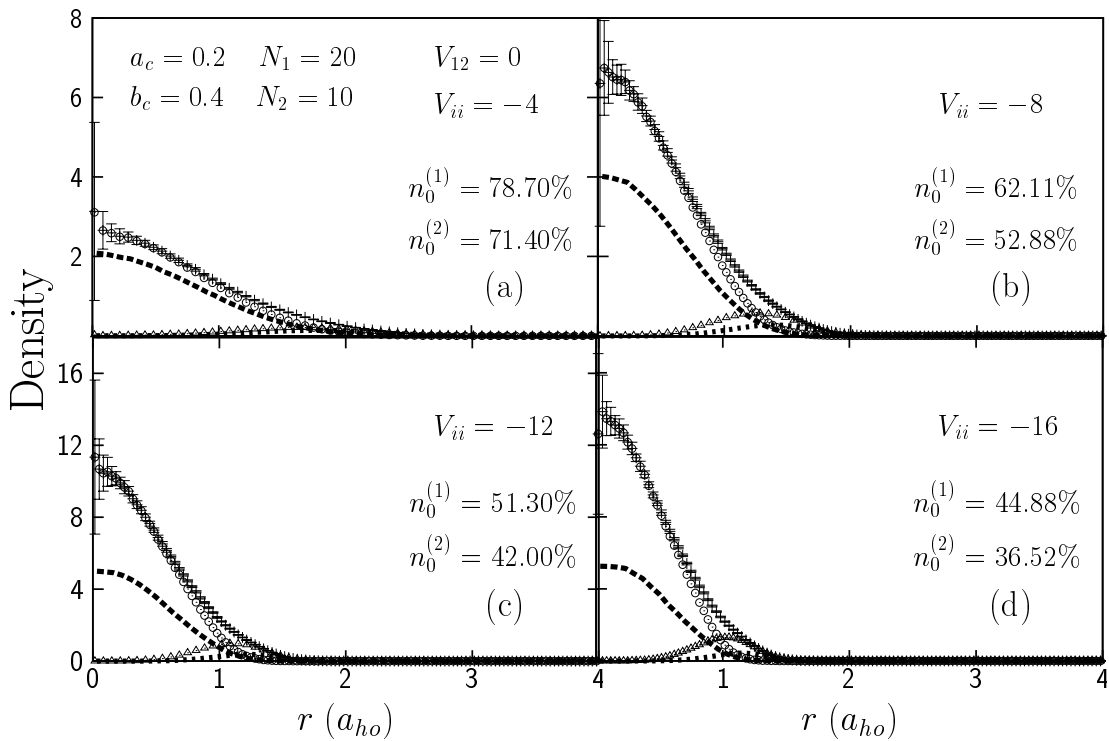}
\caption{\footnotesize\baselineskip 0.2cm As in 
Fig.\ft\ref{fig:plot.density.figures.complete.mix.stackI} but with $a_c=0.2$, 
$b_c=0.4$ and HCSW intra- and HC interspecies interactions. The intraspecies 
HCSW depth is $V_{ii}$ ($i=1,2$).} 
\label{fig:plot.densityprofiles.with.attrctve.intra.8FIGSSTACKB}
\end{figure*}

\begin{figure}[h!]
\hspace{-0.2cm}\includegraphics[width=8.8cm,bb=180 460 528 706,clip]{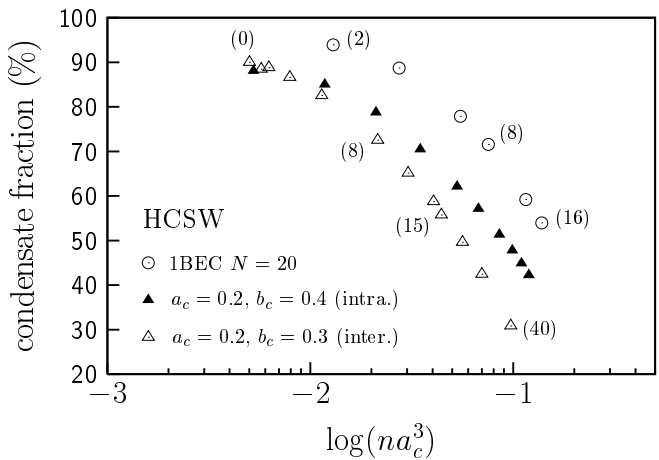}
\caption{\footnotesize\baselineskip 0.2cm Condensate fraction \cn01\ for HCSW 
2BECs compared to a {\it reference}. Open circles: ({\it reference}) HCSW 1BEC 
with $N=20$ particles and $a_c=0.2$ in which the HCSW depth $V$ is varied 
between $2$ and $16$. Solid triangles: 2BEC of 
Fig.\ft\ref{fig:plot.densityprofiles.with.attrctve.intra.8FIGSSTACKB} with 
attractive intraspecies interactions. Open triangles: 2BEC of 
Fig.\ft\ref{fig:plot.density.figures.complete.mix.stackI} with attractive 
interspecies interactions. The numbers between brackets near to some of the 
points show $|V_{ij}|$ for the corresponding systems. The points are larger than the 
error bars.} \label{fig:plot.n01andn02.for.intra.and.inter.HCSW.2BECs}
\end{figure}

\subsection{Attractive intra- and repulsive interspecies interactions}
\label{sec:attr.intra.and.rep.inter}

\subsubsection{Density profiles}

\hs Figure \ref{fig:plot.densityprofiles.with.attrctve.intra.8FIGSSTACKB} 
displays density profiles as in 
Fig.\ft\ref{fig:plot.density.figures.complete.mix.stackI} but with $a_c=0.2$ and 
$b_c=0.4$ with HCSW intra- and HC interspecies interactions ($V_{12}=0$). We keep 
$a_c$ and $b_c$ fixed and vary the intraspecies HCSW depths ($V_{ii}$, 
($i=1,2$)) in the range $V_{ii}=-4$ to $-16$ keeping $V_{11}=V_{22}$. As a 
result of the attractive intraspecies interactions, the core and the shell 
contract in volume and the density of the system grows substantially in response 
from $n_T(0)\sim 2.5$ in frame (a) to $\sim 14$ in (d) as the radius of the 
cloud shrinks from $\sim 3$ to $1.5$. The shell is pushed radially inwards 
towards the center of the trap by the confining forces of the trap. Contrary to 
Fig.\ft\ref{fig:plot.density.figures.complete.mix.stackI}, the $n_{0,1}(r)$ and 
$n_{0,2}(r)$ profiles keep following the shape of their corresponding $n_1(r)$ 
and $n_2(r)$.

\subsubsection{Condensate fractions}

\hs In what follows we investigate the condensate properties of 2BECs, this time 
with attractive intra- and repulsive interspecies interactions. This is somewhat 
the opposite case of Sec. \ref{sec:attr.inter.rep.intra} where attractive 
inter- and repulsive intraspecies interactions are used. We chiefly aim at 
revealing the difference in the results when using different types of 
combinations of repulsive and attractive interactions. Figure 
\ref{fig:plot.n01andn02.for.intra.and.inter.HCSW.2BECs} compares now the 
condensate fraction \cn01\ of the 2BEC of 
Fig.\ft\ref{fig:plot.densityprofiles.with.attrctve.intra.8FIGSSTACKB} (solid 
triangles) with HCSW intraspecies interactions (intra.) against \cn01\ of the 
2BEC of Fig.\ft\ref{fig:plot.density.figures.complete.mix.stackI} 
(open triangles) with HCSW interspecies interactions (inter.). The open circles 
display the condensate fraction $n_0$ of a HCSW 1BEC of $N=20$ particles and 
$a_c=0.2$ in a separate trap of its own, this is our {\it reference system}. Here 
the HCSW depth is varied in the range ($V=0$ to $-16$). 

\hs The condensate fraction \cn01\ of the mixture with attractive interspecies 
interactions (inter.) shows the largest depletion, although the bosons in its 
shell have a smaller HC diameter than those in the mixture with attractive 
intraspecies interactions (intra.). It seems that the attractive intraspecies 
interactions boost the value of the condensate (solid triangles) beyond the HC 
intraspecies result (open triangles). It will be shown in Sec. 
\ref{sec:rep.inter.and.inter.} below that for a number of 
purely repulsive mixtures with the same bosonic HC diameters in the cores but 
different HC diameters in the shells, the condensate depletion is larger in the 
2BECs with larger bosons in the shell. Our {\it reference} shows again the 
smallest condensate depletion which is again a manifestation of the fact that 
mixing enhances the condensate depletion beyond the 1BEC result.

\hs Substantial depletion is observed in 
Fig.\ref{fig:plot.densityprofiles.with.attrctve.intra.8FIGSSTACKB} at 
$V_{ii}=-16$ ($\sim 50\%$ for \cn01\ and $\sim 60\%$ for \dn02). If compared to 
Fig.\ft\ref{fig:plot.density.figures.complete.mix.stackI} we can see that this 
amount of depletion sets in there at $V_{12}=-15$ (frame e), that is at a 
comparable (interspecies) HCSW depth. However, the density in the latter 
($\sim 11$) is lower than the former ($\sim 14$) because the number of 
attracting pairs of two-species bosons is smaller in 
Fig.\ft\ref{fig:plot.density.figures.complete.mix.stackI} than in our case here 
which leads to a slower rise in the density with HCSW depth. 

\subsection{Repulsive inter- and intraspecies 
interactions} \label{sec:rep.inter.and.inter.}

\subsubsection{Density profiles}

\begin{figure*}[t!]
\includegraphics[width=15cm,bb = 81 147 532 707,clip]{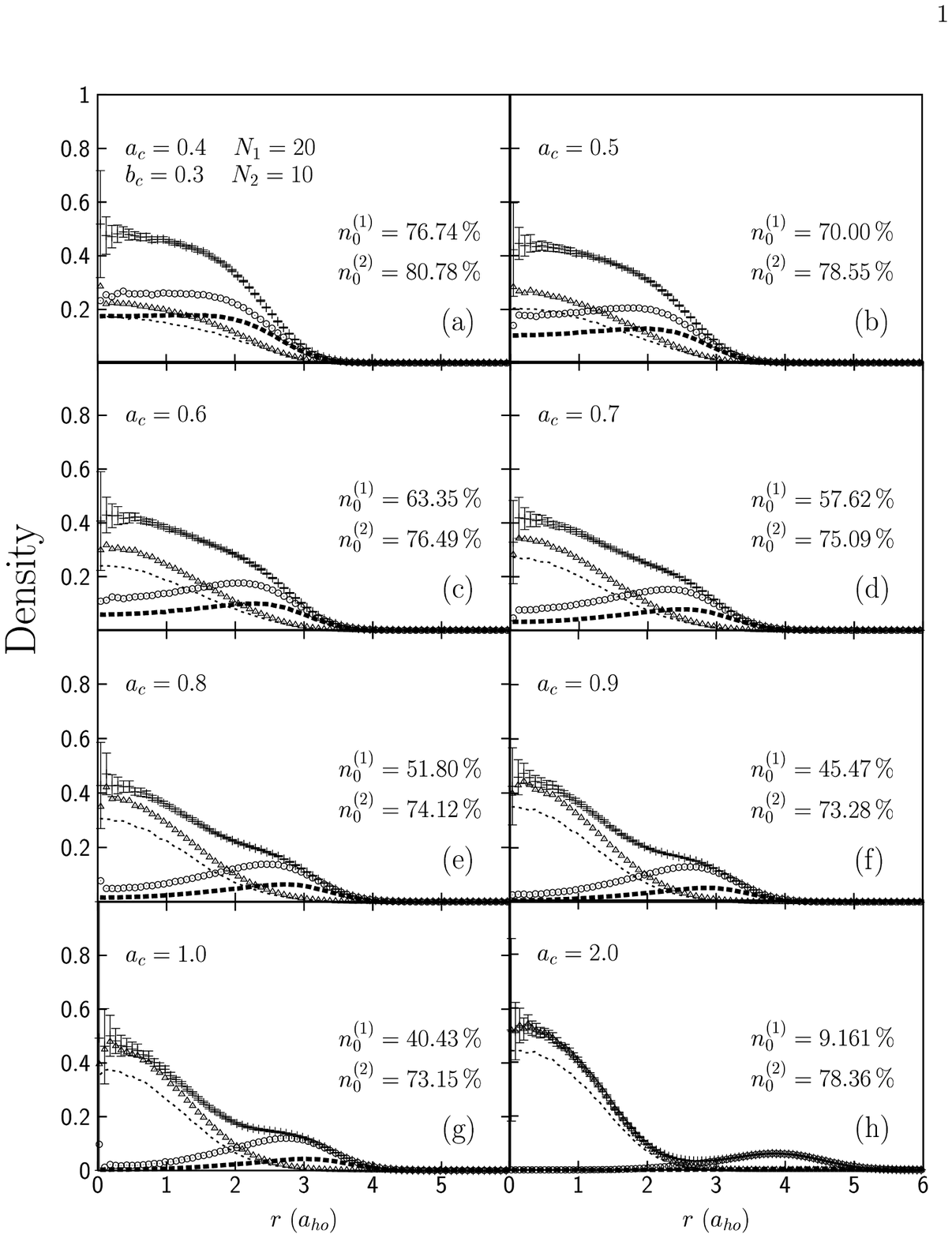}
\caption{\footnotesize\baselineskip 0.2cm As in 
Fig.\ft\ref{fig:plot.density.figures.complete.mix.stackI} but with HC 
interactions only and varying $a_c$. $b_c$ is held fixed at 0.3.\vspace{0.0cm}} 
\label{fig:plot.density.figures.HC.mix.stackI}
\end{figure*}

\begin{figure*}[t!]
\includegraphics[width=15cm,bb = 81 147 532 707,clip]{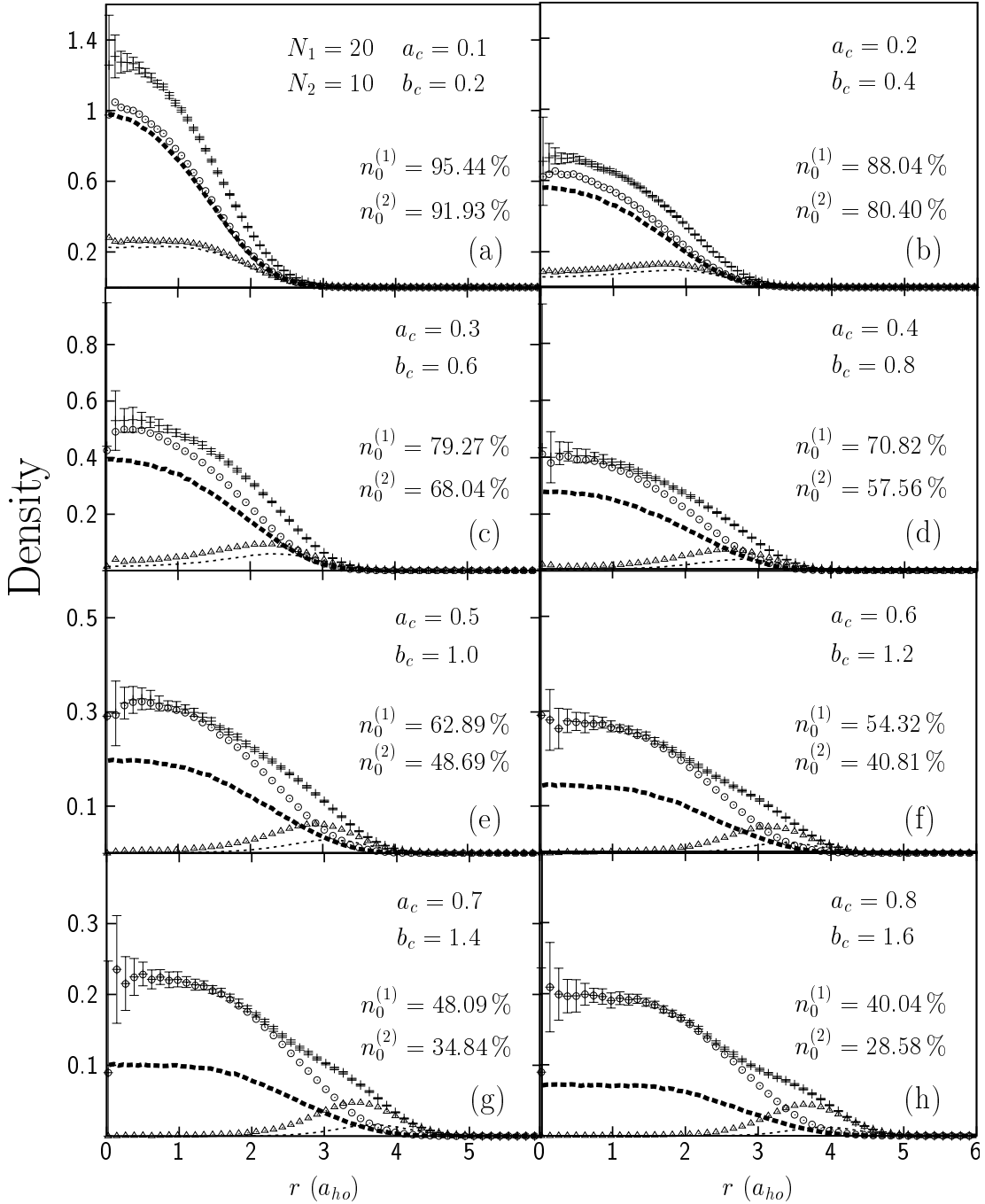}
\caption{\footnotesize\baselineskip 0.2cm As in 
Fig.\ft\ref{fig:plot.density.figures.HC.mix.stackI} but with the HC diameters 
both varying at a fixed ratio of $a_c:b_c=1:2$.} 
\label{fig:plot.density.figures.HC.mix.acandbcvary.stackIII}
\end{figure*}

\begin{figure}[t!]
\hspace{-0.3cm}\includegraphics[width=8.9cm,bb=191 457 553 709,clip]{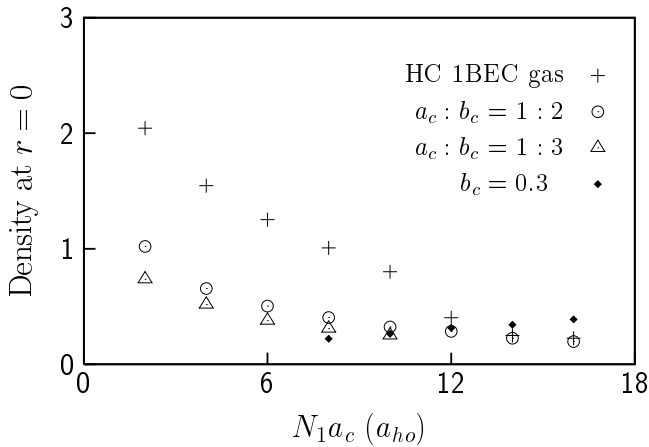}
\caption{Density at the center of the trap ($r=0$) versus 
the interaction parameter $N_1a_c$ for HC systems only. Crosses: 1BEC with 
$N=20$ particles ({\it reference}), open circles: 2BEC of 
Fig.\ft\ref{fig:plot.density.figures.HC.mix.acandbcvary.stackIII} with 
$a_c:b_c=1:2$, open triangles: 2BEC with $a_c:b_c=1:3$, solid diamonds: 2BEC 
with $b_c$ fixed and $a_c$ varying 
(Fig.\ft\ref{fig:plot.density.figures.HC.mix.stackI}). The points are larger 
than the error bars.} 
\label{fig:plot.compare.n.vs.ac.for1BEC.HC.Bose.gas.with.Fig.5}
\end{figure}

\begin{figure}[t!]
\hspace{-0.3cm}\includegraphics[width=8.9cm,bb=180 457 530 702,clip]{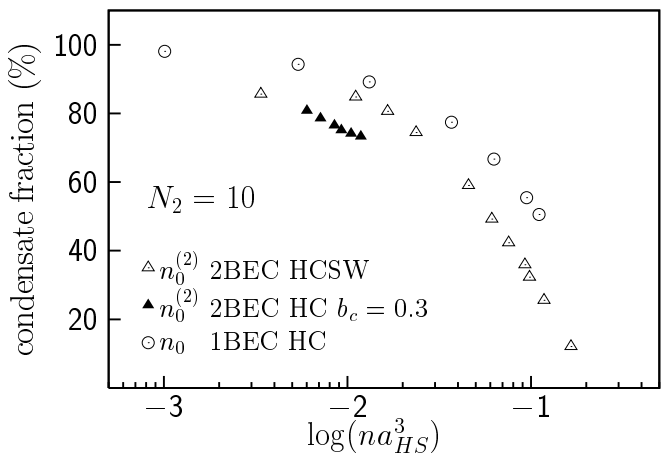}
\caption{\footnotesize\baselineskip 0.15cm Condensate 
fraction versus \xna3\ for a HC 1BEC and component 2 in 2BECs. Open triangles: 
HCSW 2BEC of Fig.\ft\ref{fig:plot.density.figures.complete.mix.stackI}, solid 
triangles: HC 2BEC of Fig.\ft\ref{fig:plot.density.figures.HC.mix.stackI}, open 
circles: HC 1BEC with $N=10$ particles ({\it reference}). The points are larger 
than the error bars.} \label{fig:plotN10mixedHCandHCSWand1BECcomparisons}
\end{figure}

\begin{figure}[t!]
\includegraphics[width=8.7cm,bb = 222 305 567 554,clip]{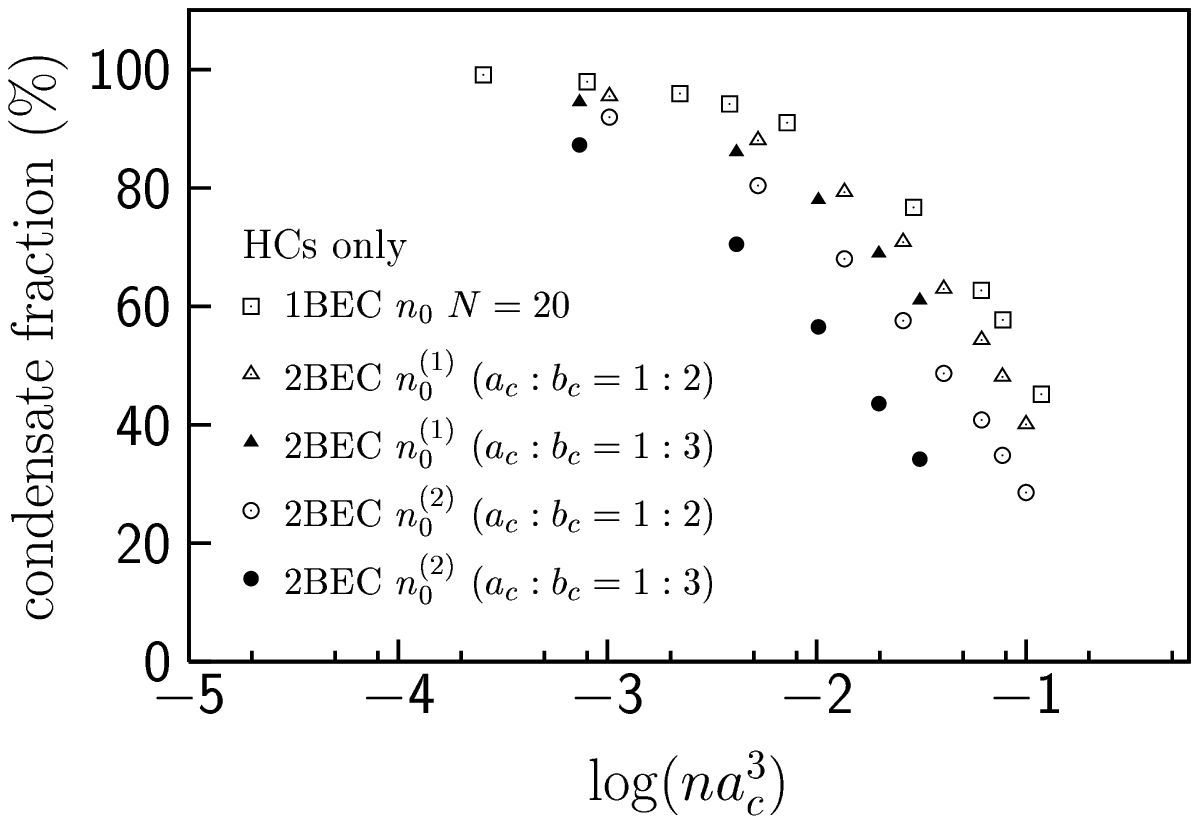}
\caption{\footnotesize\baselineskip 0.2cm Condensate fraction versus \xna3\ for 
HC systems only. Open squares: 1BEC of $N=20$ particles, open triangles and 
circles: \cn01\ and \dn02\ for the 2BEC of 
Fig.\ft\ref{fig:plot.density.figures.HC.mix.acandbcvary.stackIII}, solid 
triangles and circles: \cn01\ and \dn02\ for a HC 2BEC with $a_c:b_c=1:3$. The 
points are larger than the error bars.} 
\label{fig:plotn0in1BECHCandn01in2BECHCcomparisons}
\end{figure}

\hs Figure \ref{fig:plot.density.figures.HC.mix.stackI} displays the VMC spacial 
density distributions of HC 2BECs with various interspecies interactions and the 
same definitions of points as in 
Fig.\ft\ref{fig:plot.density.figures.complete.mix.stackI}. Here $a_c$ is varied 
while $b_c$ is kept fixed at 0.3. We can see that mixing of the two components 
is enabled up to $a_c=1.0$ before complete phase-separation sets in. On 
increasing $a_c$ beyond $b_c$ in frames (a-h), component 1 (of HC $a_c$) is 
gradually pushed out towards the edges of the trap with the rise of interspecies 
repulsion $(a_c+b_c)/2$. This is contrary to our expectations because we thought 
that component 2 (of HC $b_c$) would be pushed out instead since it is the 
lighter of the two given that $m_1/m_2=1.200$ and $\omega_1=\omega_2$. This can 
be explained as follows. Essentially, as the HC diameter of the bosons of one 
component increases, the Bose gas expands in size in order to accomodate the 
larger bosons. As a result, the component with lower intraspecies repulsion 
``falls" into the center of the trap seeking the minimizaton of the total 
repulsive potential energy. As the HC diameter $a_c$ is increased, the 
interspecies repulsion rises pressurizing the core radially towards the center 
of the trap. As a result, the density of the core $n_2(0)$ rises by a factor of 
$\approx 2$ from frame (a) to (h) and nearly as from frame (d) on, $n_1(0)$ 
begins to approach zero and the two components begin to separate into a shell 
and a core. If we imagined removing the shell completely from the trap, the core 
will expand and become almost uniform in density and therefore ``flat" in shape 
\cite{DuBois:01,Sakhel:02}. We found that it is very hard to ``squeeze" the core 
further to higher density by increasing $a_c$ beyond 2. In frame (h) total phase 
separation has occured leaving a dip at the boundary between the two components. 
One could imagine placing a third-species particle in that dip as it is a 
potential trap by itself. Contrary to the case of 
Figs.\ft\ref{fig:plot.density.figures.complete.mix.stackI}(d-f), the condensate 
density distributions of both components follow the shape of their total 
densities up to phase separation.

\hs With repulsive interactions only in these mixtures, we can always have 
stable systems if there is sufficient repulsion between the bosons of the core 
counteracting the outer pressure arising from the shell. Otherwise a dilute 
core collapses readily under the heavy pressure of a dense shell. 

\hs Figure \ref{fig:plot.density.figures.HC.mix.acandbcvary.stackIII} displays 
the MC density distributions of a 2BEC with HC interactions only where $a_c$ and 
$b_c$ are both increased at a fixed ratio $a_c:b_c=1:2$. In this case the 
density of the core $n_1(0)$ decreases because both $a_c$ and $b_c$ are 
increased. A peculiar result is that even at very large values of the 
interspecies repulsions no complete phase separation is observed as it occurs in 
Fig.\ft\ref{fig:plot.density.figures.HC.mix.stackI}h. Some uniformity in the 
density distribution of the core arises at the larger $a_c$.

\hs In Fig.\ft\ref{fig:plot.compare.n.vs.ac.for1BEC.HC.Bose.gas.with.Fig.5} we 
make comparisons between densities at the center of the trap as a function of 
the interaction parameter $N_1a_c$ for various systems with repulsive 
interactions. 
The crosses are for a HC 1BEC of 20 particles {\it in a separate trap of its 
own} and the same trap length as before $a_{ho}=\sqrt{\hbar/m_1\omega_1}$. The 
open circles and triangles are, respectively, for the core in 
Fig.\ft\ref{fig:plot.density.figures.HC.mix.acandbcvary.stackIII} and an 
additional mixture with $a_c:b_c=1:3$ whose density profiles we do not reveal. 
The solid diamonds are for the core in 
Fig.\ft\ref{fig:plot.density.figures.HC.mix.stackI}. Thus the goal is to show 
the effect of mixing a HC 1BEC with various other 1BECs on the central core 
density of the system. The density $n(0)$ of the HC 1BEC drops as it is mixed 
with another component, and for a larger ratio of $b_c$ relative to $a_c$ the
core density drops further. The density of the core with one of the HCs fixed 
($b_c=0.3$) varies only slightly as $a_c$ increases. Note that the values of 
the core density at $r=0$, except for the latter case, converge at the higher 
$N_1a_c$.

\subsubsection{Condensate fraction}

\hs Figure \ref{fig:plotN10mixedHCandHCSWand1BECcomparisons} compares the 
condensate fraction \dn02\ ofFig.\ft\ref{fig:plot.density.figures.HC.mix.stackI} 
($a_c:b_c=1:2$) to two other systems. The open circles display $n_0$ for a HC 
1BEC of 10 particles {\it in a separate trap of its own} and the same trap 
length as before $a_{ho}=\sqrt{\hbar/m_1\omega_1}$. The solid triangles display 
\dn02\ for the HC 2BEC of Fig.\ft\ref{fig:plot.density.figures.HC.mix.stackI} 
with $b_c$ fixed and $a_c$ varying. The open triangles display \dn02\ for the 
HCSW 2BEC of Fig.\ref{fig:plot.density.figures.complete.mix.stackI} with $a_c$ 
and $b_c$ fixed and $V_{12}$ varying. We note that the condensate depletion is 
highest in the HC 2BEC of Fig.\ft\ref{fig:plot.density.figures.HC.mix.stackI}. 
The depletion of the condensate in the HCSW 2BEC is less pronounced. This 
reveals that the repulsive interspecies interactions play a more pronounced role 
in depleting the condensates of the mixture than attractive interspecies 
interactions. The attractive interspecies interactions boost the condensate 
somewhat above the HC-interspecies interactions result. The depletion of the 
condensate is lowest for a 1BEC. Thus the mixing of condensates enhances their 
depletion due to the presence of interspecies interactions of various strengths 
as compared to the case when they are separate, each in a trap of its own. 

\hs Figure \ref{fig:plotn0in1BECHCandn01in2BECHCcomparisons} displays chiefly 
the condensate fractions of the components as a function of \nac3\ at the center 
of the trap in various HC systems compared to a {\it reference}. The open 
squares represent the condensate fraction $n_0$ of a HC 1BEC of 20 particles 
({\it reference}). The open triangles represent the condensate fraction \cn01\ 
of the 2BEC of Fig.\ft\ref{fig:plot.density.figures.HC.mix.acandbcvary.stackIII} 
where $a_c$ and $b_c$ are varied at a constant ratio of 1:2 and the solid 
triangles that at a ratio of 1:3, respectively. In addition, and for 
further comparison, the open and solid circles represent \dn02\ of the latter 
two mixtures, respectively. We can see that \cn01\ is lower for $a_c:b_c=1:3$ 
than for $1:2$ and the same is true for \dn02. The latter values of \cn01\ are 
lower than those of the HC 1BEC displayed for comparison. This shows again that 
mixing and a larger interspecies interaction enhance the depletion of the 
condensates in each component beyond the 1BEC result.

\subsection{Energies}

\hs In this section we compare our VMC energies for HC 2BECs against the 
energies calculated by an approximate model derived from mean-field results in 
Appendix A. The estimate that we obtained for the total energy of the mixture 
$E^{MFA}$ is given by Eq.\ft(\ref{eq:E_MFA}) where MFA stands for mean-field 
approximation.

\begin{figure}[t!]
\includegraphics[width=8.8cm,bb = 192 460 554 707,clip]{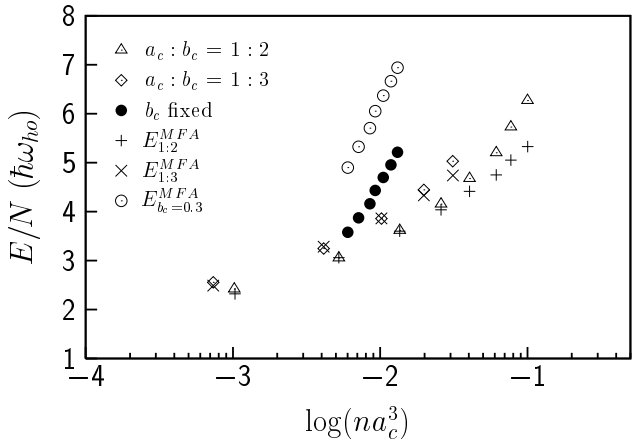}
\caption{\footnotesize\baselineskip 0.2cm VMC energies and energies of 
Eq.(\ref{eq:E_MFA}) ($E^{MFA}/N$) versus the HC density \xna3. Open triangles: 
mixture in Fig.\ft\ref{fig:plot.density.figures.HC.mix.acandbcvary.stackIII} 
($a_c:b_c=1:2$), open diamonds: the mixture with $a_c:b_c=1:3$, and solid 
circles: mixture in Fig.\ft\ref{fig:plot.density.figures.HC.mix.stackI} 
($b_c=0.3$). Crosses, times, and open circles: corresponding estimates $E^{MFA}/N$ 
in the same order. The points are larger than the error bars.} 
\label{fig:plotvmcenergiesvsnac3}
\end{figure}

\hs Figure \ref{fig:plotvmcenergiesvsnac3} displays our 
VMC energies for the repulsive mixtures investigated in 
Figs.\ft\ref{fig:plot.density.figures.HC.mix.acandbcvary.stackIII} with 
$a_c:b_c=1:2$ (open triangles) and \ref{fig:plot.density.figures.HC.mix.stackI} 
with $b_c=0.3$ (solid circles). The open diamonds show the additional VMC 
calculations for HC 2BECs with $a_c:b_c=1:3$ similar to 
Fig.\ft\ref{fig:plot.density.figures.HC.mix.acandbcvary.stackIII}. The crosses, 
times, and open circles display $E^{MFA}$ for the systems indicated by the open 
triangles, diamonds, and solid circles, respectively. We note that there is good 
agreement between the energies $E_{1:2}^{MFA}$ and the VMC results at the lower 
\nac3\ but then they begin to diverge somewhat at the higher densities. The same 
is true for $E_{1:3}^{MFA}$. $E_{b_c=0.3}^{MFA}$ largely do not agree with the 
VMC results but show the same trend in their values. This might be chiefly due 
to the fact that the TF radius of the core is not a good representation of the 
cloud radius for this particular case because the two components completely 
phase-separate at the higher interspecies repulsions. In all cases, the energies 
rise with the HC densities. The rise is steepest when $b_c$ is fixed and $a_c$ 
varied, the reason being due to the fact that the core density varies slowly 
with the rise of $a_c$ (see 
Fig.\ft\ref{fig:plot.density.figures.HC.mix.stackI}). That is as the shell is 
expelled towards the edges of the trap, the potential energy rises faster than 
the change in $n_2(0)b_c^3$ compared to the other systems.

\subsection{Effect of mass ratio}

\hs In this section we present the role of the mass ratio $m_{ratio}=m_1/m_2$ in 
determining the properties of mixed Bose gases. We consider two mass-ratios, the 
previous $m_{ratio}=1.2$ and a new $m_{ratio}=5$ and in order to keep the trap 
length unchanged, we only change $m_2$. We further note that changing 
$m_{ratio}$ changes the energy of the system since $m_1/m_2$ and $m_2/m_1$ 
appear explicitly in the Hamiltonian (\ref{eq:Hamiltonian2}).

\begin{figure}[t!]
\hspace{-0.2cm}\includegraphics[width=8.8cm,bb=235 457 587 703,clip]{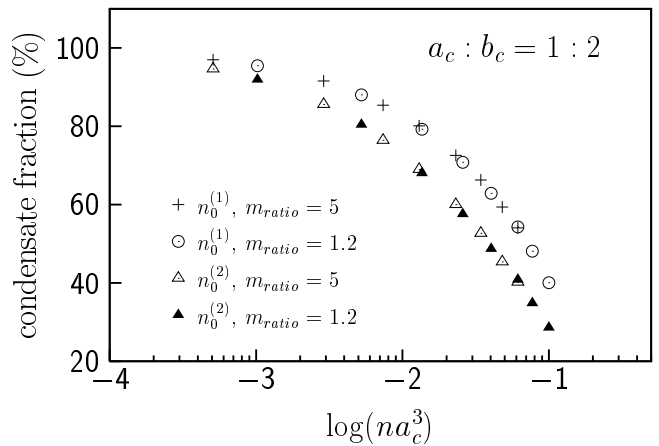}
\caption{\footnotesize\baselineskip 0.2cm Condensate 
fractions versus \xna3\ for two mass ratios of a HC 2BEC with $a_c:b_c=1:2$. 
Crosses and open circles: \cn01\ for $m_{ratio}=5$ and 1.2, respectively, open 
and solid triangles: \dn02. The points are larger than the error bars.} 
\label{fig:plotn0vsnac3formratio1.2and5}
\end{figure}

\begin{figure}[t!]
\hspace{-0.2cm}\includegraphics[width=8.8cm,bb = 235 457 589 703,clip]
{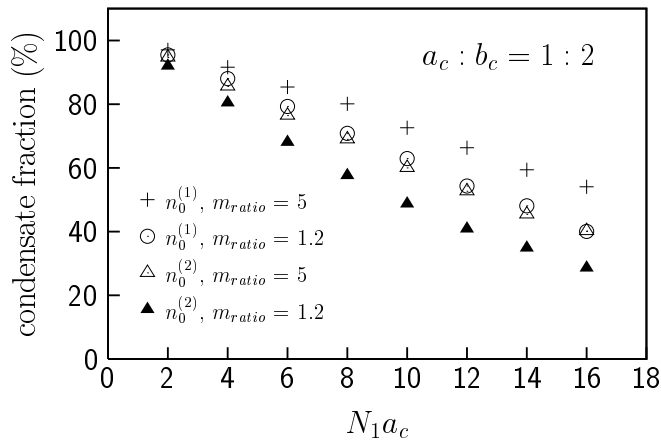}
\caption{\footnotesize\baselineskip 0.2 cm As in 
Fig.\ft\ref{fig:plotn0vsnac3formratio1.2and5} but versus $N_1a_c$ instead of 
\xna3. The points are larger than the error bars.} 
\label{fig:plotn0vsNacformratio1.2and5}
\end{figure}

\begin{figure}[t!]
\hspace{-0.2cm}\includegraphics[width=8.8cm,bb = 180 430 567 676,clip]
{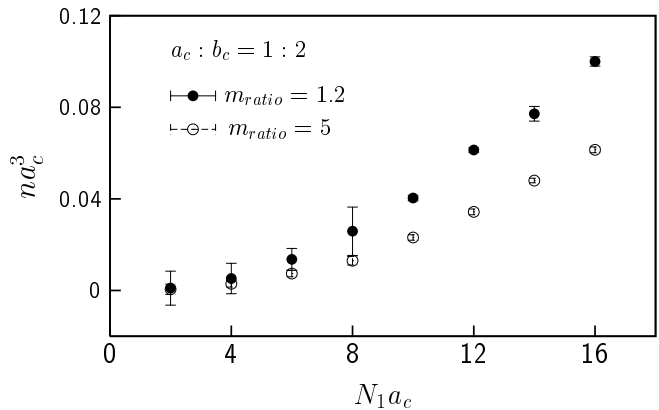}
\caption{\footnotesize\baselineskip 0.2 cm $na_c^3$ as a 
function of the interaction parameter $N_1a_c$ for two mass-ratios of a HC 2BEC 
with $a_c:b_c=1:2$.} \label{fig:plotna3vsNaformaratios5and1.2}
\end{figure}

\begin{figure}[t!]
\includegraphics[width=8.8cm,bb = 192 458 548 706,clip]{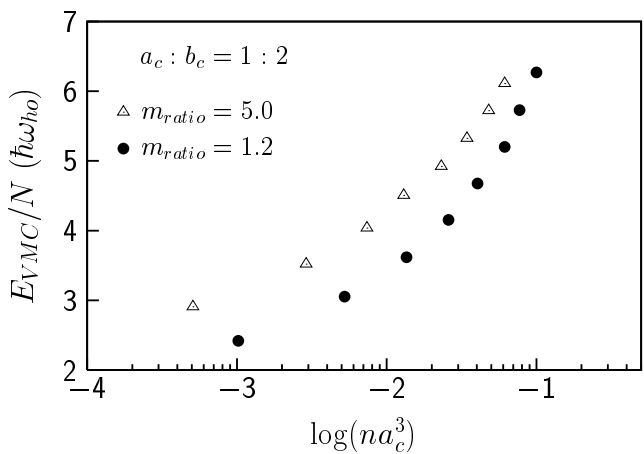}
\caption{\footnotesize\baselineskip 0.2 cm VMC energy for the same systems in 
Fig.\ft\ref{fig:plotna3vsNaformaratios5and1.2} and two mass ratios. The points 
are larger than the error bars.} 
\label{fig:plot.compare.E.vs.nac3.for.Fig.5.w2massr}
\end{figure}

\hs We consider the systems depicted in 
Fig.\ft\ft\ref{fig:plot.density.figures.HC.mix.acandbcvary.stackIII} with 
$m_{ratio}=1.2$ and compare its properties with those of exactly these same 
systems evaluated at $m_{ratio}=5.0$. Figure 
\ref{fig:plotn0vsnac3formratio1.2and5} displays the condensate fractions of the 
latter systems vs \nac3\ where the condensate fraction \cn01\ (and \dn02)
is the same for both values of $m_{ratio}$. Therefore $m_{ratio}$ has no 
influence on the relation between condensate fraction and HC density. The 
scenario is however different if one plots the condensate fractions as a 
function of the HC interaction parameter $N_1a_c$ as in 
Fig.\ft\ref{fig:plotn0vsNacformratio1.2and5}. The crosses and open circles show 
\cn01\ and the open and solid triangles \dn02\ each for $m_{ratio}=5$ and 1.2, 
respectively. We can see that the condensate fractions for $m_{ratio}=5$ are 
higher than for 1.2 because the central HC densities are lower for 
5. Effectively, as $m_2$ is reduced to increase $m_{ratio}$, the trapping forces 
confining the shell ($-\nabla \frac{1}{2} m_2 \omega^2 r_2^2$) are reduced 
accordingly. Thus, the cloud of the mixture expands as the pressure of the shell 
on the core is lifted causing the central densities to decline in favor of an 
increase of the condensate fractions. Figure 
\ref{fig:plotna3vsNaformaratios5and1.2} displays \nac3\ as a function of 
$N_1a_c$ for the latter systems where \nac3\ for $m_{ratio}=5$ is lower than for 
1.2 at higher $N_1a_c$ as explained above. As we increase $m_{ratio}$, the 
energy of the system rises as demonstrated in 
Fig.\ft\ref{fig:plot.compare.E.vs.nac3.for.Fig.5.w2massr} where $E_{VMC}/N$ is 
plotted against \nac3\ for two $m_{ratio}$ values and $a_c:b_c=1:2$.

\section{Discussion}\label{sec:discussion}

\hs We discuss now important details on the results of our calculations and 
connect to the previous literature. We elaborate on the mixing and phase 
separation of components, stability of the mixtures and the origins of the 
enhanced condensate depletion. First of all, however, we mention briefly the 
work of Ma and Pang \cite{Ma:06} which is most relevant to ours.

\subsection{The work of Ma and Pang}

\hs Ma and Pang \cite{Ma:06} investigated HC 2BECs trapped in a 
three-dimensional isotropic trap at finite temperature using path-integral 
quantum Monte Carlo simulations. Their main interest was in the structure of the 
mixtures, i.e., the densities and their profiles. They particularly concentrated 
on the conditions under which phase separation occurs and treated the 
two-species system as two subsystems each of which contains one species with its 
own statistics. However, the two species do not behave independently of 
each other due to the interspecies interactions. They found that by changing the 
mass ratio of the components $m_2/m_1$ the lighter particles are pushed outward 
and form a shell surrounding the heavier core. Further, the density and 
condensate fractions of the mixture drop with a rise of the interspecies 
interactions. When identical external potentials are used, no phase separation 
is observed, but when they are different phase separation occurs. They also 
found that the spacial phase separation is independent of $m_2/m_1$ and that 
the species with the larger scattering length favors the formation of a 
low-density outer shell. 

\hs Now in our work here we conducted our calculations at zero temperature and 
we simulated two-species Bose gases with both attractive and repulsive 
interactions. We particularly concentrated on the role of the inter- and 
intraspecies interactions on the enhancement of the condensate depletions in the 
mixture as compared to the case when each component is in a separate trap of its 
own. We followed Ma and Pang in treating each component as a subsystem with its 
own properties. In addition to their investigations on the effect of the mass 
ratio, we additionally investigated its effect on the energies and condensate 
fractions. 

\subsection{Mixing and demixing}

\hs In the case of intercomponent attraction as in 
Fig.\ft\ref{fig:plot.density.figures.complete.mix.stackI} no phase 
separation occurs because the two gases attract each other. Note that the 
attractive interspecies interactions at some point overwhelm the repulsive 
intraspecies interactions as identified by the large increase in the overall 
density $n_T(r)$ of the system. That is, the repulsive intraspecies interactions 
do not become attractive, they just get overwhelmed similar to a case discussed 
by Chui and Ryzhov \cite{Chui:04}. Further, the two gases are now trapped by the 
attractive potential of each other such that the importance of the external trap 
is underplayed. Since in all our calculations the minima of the two confining 
potentials coincide, the two components interpenetrate completely 
\cite{Cornell:98} and are drawn together into what resembles a 1BEC system 
acting similarly to it. We conclude that in this case, the attractive 
interspecies interactions play a more pronounced role than the repulsive ones in 
determining the properties of the systems. Therefore the condensates are able to 
migrate towards the edges of the trap at the higher densities \xna3\ as in the 
case of a HC 1BEC \cite{DuBois:01}. In the somewhat opposite case of 
Fig.\ft\ref{fig:plot.densityprofiles.with.attrctve.intra.8FIGSSTACKB} the 
intercomponent repulsion pushes the condensate of the core towards the center of 
the trap.

\hs In the case of Figs.\ft\ref{fig:plot.density.figures.HC.mix.stackI} and
\ref{fig:plot.density.figures.HC.mix.acandbcvary.stackIII}, we find chiefly that 
full mixing is impossible at large repulsive interspecies interactions. Larger 
interspecies repulsion leads to full phase separation as in 
Fig.\ft\ref{fig:plot.density.figures.HC.mix.stackI}h and even though the shell 
there has a much lower density than the core, substantial depletion $\sim 20\%$ 
is still observed which is attributed to the presence of large bosons in the 
shell. Note that it is hard to define a density \xna3\ for the shell as it is 
expelled towards the edges of the trap.

\hs An investigation of the detailed nature of the overlap region between the 
shell and core is also important since it influences properties such as the 
ground state energy, the excitation spectrum and the collisional relaxation 
rates as outlined earlier by Barankov \cite{Barankov:02}. He explored the 
boundary between two repulsively interacting condensates in the weak and strong 
separation limits and found that the asymptotic behaviour of each condensate far 
from the boundary is determined by its correlation (healing) length. In the case 
of strong separation, he found that there exists a hollow in the total density 
profile which is very deep. The latter allows the investigation of one-particle 
excitations at the boundary between the components as well as surface wave 
excitations due to the surface tension \cite{Barankov:02}. As a result of the 
full separation in Fig.\ft\ref{fig:plot.density.figures.HC.mix.stackI}, we also 
observe a hollow in the total density profile between shell and core as 
discussed by Barankov. In the future one could add {\it one} foreign particle to 
be trapped by this hollow and investigate its energy as a function of some 
property of the mixture using the Monte Carlo method. Such a hollow is however 
not observed in 
Fig.\ft\ref{fig:plot.density.figures.HC.mix.acandbcvary.stackIII} because no 
complete phase-separation occurs. The reason is because as the HC diameters of 
the bosons in the core are increased, the bosons spread out and the core 
expands. As a result, these bosons penetrate into the shell which is pushed in 
the opposite direction towards the center of the trap by the confining forces of 
the external potential. This is however not the case in 
Fig.\ft\ref{fig:plot.density.figures.HC.mix.stackI} as the HC diameters of the 
bosons in the core are kept fixed and that of the shell increased. Thus complete 
phase-separation is only possible when the size of the bosons in only one 
component is increased. This has also been confirmed by Ma and Pang previously.

\hs Two length scales can be used to characterize a two-component BEC 
\cite{Trippenbach:00}: one is the penetration depth, the other is the healing 
length. The penetration depth is a measure for the width of the overlap region 
and, as we can see from above, a function of the interspecies interactions. It 
is largest in the case of attraction between the two components as it is equal 
to the radius of the cloud whereas in the case of large intercomponent repulsion 
it is smaller than the radius of the cloud. The penetration depth is reduced as 
the intercomponent repulsion rises. Nevertheless, complete mixing is 
still possible at moderate repulsive interspecies forces as demonstrated in 
Figs.\ft\ref{fig:plot.density.figures.HC.mix.stackI} and 
\ref{fig:plot.density.figures.HC.mix.acandbcvary.stackIII}. Since the density 
profile of the core in the latter two systems is very much influenced by the 
presence of a shell, we anticipate that the healing length of a 1BEC changes 
upon mixing with a shell. In what follows, we discuss some of the previous 
literature in connection to our current observations.

\hs Shchesnovich \ea\ \cite{Shchesnovich:04} studied a \rb85\ and \Rb87\ 2BEC by 
varying the interspecies interactions. They found that these two components 
would not separate if the interspecies interactions are attractive and in this 
paper we have verified this point as well. They argued that a separation of 
the two species takes place when the energy gain due to the attractive 
intraspecies interactions overwhelms the quantum pressure at the interface of 
the two species. 

\hs Cornell \ea\ \cite{Cornell:98} reviewed some early results on mixed 
condensates and provided a qualitative exegesis of the theoretical and 
experimental techniques that are involved. They found that there is a critical 
value for the interaction term $a_{12}^c = \sqrt{a_1 a_2}$ beyond which phase 
separation occurs with little spatial overlap. This is when the scattering 
length $a_1$ of component 1 becomes larger than $a_2$ of component 2, causing 
atoms 1 to move favorably towards the edges of the cloud forming a spherical 
shell around the core consisting of atoms 2. Our results are in line with those
of Cornell \ea\ \cite{Cornell:98} and also Hall \ea\ \cite{Hall:98} as we also 
observe that the component with the larger bosonic hard core diameter migrates 
to the edges of the trap. 

\hs Shi \ea\ \cite{Shi:00} studied the phase separation of two-species trapped 
and untrapped Bose gases at finite temperature and found that the interspecies 
interactions affect the formation and depletion of the two condensates and lead 
to spatial phase-separation of the mixture. They argued that the shell is 
trapped in an effective potential which has a minimum away from the center of 
the trap close to the surface of the core. This effective trap is a combination 
of the traps confining the mixture and the interspecies interactions. According 
to Shi \ea\ then, condensation of the shell happens at the surface of the core 
and indeed we do observe a condensate in the shell as displayed in 
Figs.\ft\ref{fig:plot.density.figures.HC.mix.stackI} and 
Figs.\ft\ref{fig:plot.density.figures.HC.mix.acandbcvary.stackIII}.

\subsection{Why do we use a HCSW?}\label{sec:whyHCSW?}

\hs Particularly the HCSW is a suitable potential to describe the attractive 
interactions between the HS bosons in this work, since it is a HC contact 
interaction plus an attractive tail added to the HC and the HC diameter is the 
same as the HS diameter of the bosons. Another reason for choosing the HCSW
is to simulate a Feshbach resonance. This is because the HCSW has a well defined 
range, width, and depth and via these parameters one can easily tune the 
scattering length to be at the Feshbach resonance using Eq.(\ref{eq:scattl}) 
below when $a\,\rightarrow\,\pm\infty$ in order to check any instabilities 
(or stabilities) arising from this. In terms of designing the trial wave 
function, the exact solution of the two-body Schr\"odinger equation 
interacting via a HCSW lead us in the construction of a flexible Jastrow 
function for HS bosons with attractive interactions as mentioned in 
Sec. \ref{sec:TrialWvfn}.

\subsection{HCSW parameters used}\label{sec:HCSW-properties}

\hs Our main purpose for the choices of the previous values of the
HCSW parameters in this work was to provide a qualitative study of the 
properties of trapped Bose-gas mixtures with attractive interactions and 
to reach a qualitative understanding of the role of the interatomic interactions 
in these properties. We first remind the reader that the values of the HC 
diameters have been chosen to enable substantial depletion of the condensate. 

\subsubsection{Range}

\hs The range of the HCSW ($R=0.59a_{ho}$) used in this work is 
of the same order of magnitude as that used by Astrakharchik \ea\ 
\cite{Astrakharchik:04} for another model potential of the form 
$V(r)\,=\,-V_0/\cosh^2(r/r_0)$. Here $r_0$ determines the range and they set 
$r_0=0.1a_\rho$ where $a_\rho$ is the transverse oscillatory trap length for
a highly elongated trap. Their $a_\rho$ is small because of tight confinement
along the transverse direction, similarly our $a_{ho}$ is also considered to 
be small since we use a tight trap.

\subsubsection{Depth}

\hs We use a shallow HCSW which is much weaker
than a realistic interatomic potential \cite{vanKempen:02,Geltman:01}, drawing 
our justification from what has been noted before by Gao \cite{Gao:03}. He
discussed improved interatomic model interactions beyond the HS potential or
delta function pseudopotential used in Gross-Pitaevskii theory. These model
potentials are simple in a sense that they are shallow and are applicable in
quantum few-body and many-body systems. 

\hs One of the most important points relevant to our work mentioned by Gao is 
that a real interatomic potential can simply become unmanagable if used in 
few-body or many-body calculations. A key conclusion in his paper is that the 
real potential in a many-body system around the threshold, such as a BEC state, 
no matter how deep this potential might be, can be replaced by an effective 
shallow potential that supports only one or two bound states. Gao shows that by 
using shallow model potentials, much weaker than the real potentials, the 
results are in good agreement with those using a real interatomic potential. 

\subsection{Artificial stability of the mixtures, Feshbach resonance, and
negative energies}\label{sec:artificial-stability}

\hs The HC potentials used in this investigation, whether attractive or 
repulsive, prevent real collapse and therefore the mixtures are always stable 
and can not collapse to a singularity. In that sense, we speak of an  
``artificial stability" \cite{Mullin:06a}. This HC is present in the HCSW 
potential since upon ``switching on" a HCSW one effectively adds to the HC 
potential an attractive tail, well defined in width and depth. Thus whatever the 
depth of the HCSW is, the bosons will not be able to approach each other to 
distances lesser than the HC diameter of the interactions ($a_c$, $b_c$, or 
$(a_c+b_c)/2$) as imposed by the Jastrow functions. As mentioned in 
Section \ref{sec:TrialWvfn}, the Jastrow function of the HCSW has a short-range 
repulsive and a long-range attractive part. The short-range part of the HCSW 
Jastrow keeps the bosons at some average distance away from each other, whereas 
the attractive part tries to bring them closer together. The balance between 
the repulsive and attractive parts, keeps the system in equilibrium. In the case 
of repulsive interactions only, the system is primarily balanced by the 
repulsive HC and the external confining potential. 

\subsubsection{At the Feshbach resonance}

\begin{figure}
\includegraphics[width=8.4cm,bb= 176 462 531 701,clip]{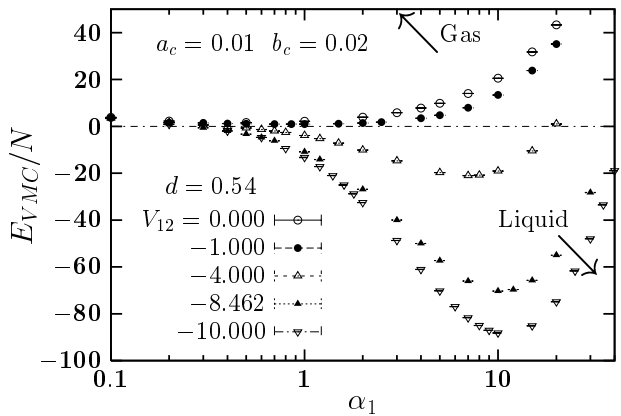} 
\caption{\footnotesize\baselineskip 0.2cm Energy vs $\alpha$ showing stabilities 
at high HCSW depths for a two-component Bose gas of $N_1=20$ and $N_2=10$ 
particles, $a_c=0.01$, $b_c=0.02$, $d=0.54$ and various interspecies 
HCSW depths $V_{12}$ shown. The first Feshbach resonance for the system is at 
$V_{12}=-8.462$} \label{fig:plotEvsalpha.R0d25.R0d54}
\end{figure}

\hs If one should increase the HCSW depths to values up to the first Feshbach 
resonance and beyond, the systems begin to shrink to very high densities. At 
this stage their energy is mainly potential (negative) and a large fraction 
of the bosons reside inside the HCSW. However they still show the artificial 
stability discussed above. For example in Fig.\ref{fig:plotEvsalpha.R0d25.R0d54} 
we demonstrate how the stability of a two-species Bose gas of $N_1=20$, $N_2=10$
particles, $d=0.54$, and HS diameters $a_c=0.01$, $b_c=0.02$, respectively, 
shifts to higher $\alpha$ values as $V_{12}$ is increased from shallow to deep 
values, even up to the first Feshbach resonance at $V_{12}=-8.462$ and beyond. 
The intraspecies interactions are repulsive HCs. The HCSW depth corresponding to 
a Feshbach resonance is obtained from the condition that 
$a\rightarrow\pm\infty$, where $a$ is the s-wave scattering length of the HCSW. 
And according to Giorgini \ea\ \cite{Giorgini:99} $a$ is given by

\begin{equation}
a\,=\,R_c + (R-R_c)\left\{1\,-\,\frac{\tan[K_0(R-R_c)]}{K_0(R-R_c)}\right\}, 
\label{eq:scattl}
\end{equation}

where $R_c$ is the HC diameter (in our case $a_c$ for component 1, $b_c$ for 
component 2, or $(a_c+b_c)/2$ for the mixture), $R$ the edge, and 
$K_0\,=\,\sqrt{V_0 m/\hbar^2}$ the wave vector of the HCSW. In trap 
units $R_c\,\rightarrow\,R_c/a_{ho}\,=\,\tilde{R_c}$ and similarly for $R$, 
$V_0\,\rightarrow\,V_0/\hbar\omega_{ho}\,=\,\tilde{V_0}$, and thus 
$K_0\,\rightarrow\,\sqrt{\tilde{V_0}\,\hbar\omega_{ho} m/\hbar^2}\,=\,
\sqrt{\tilde{V_0}}/a_{ho}\,=\,\tilde{K_0}$. 

\hs In Fig.\ref{fig:plotEvsalpha.R0d25.R0d54} at $V_{12}=-10$ the wave function 
is very much contracted and its density has risen substantially as indicated by 
a large Gaussian variational parameter $\sim 10$. Note that one still 
obtains a deep negative energy minimum at the first Feshbach resonance 
manifesting the strong stabilizing factor of the HCs. Thus even if we 
reach the Feshbach resonance and surpass it while varying the HCSW depth, the 
systems remain artificially stable. Going back to the previous 
Fig.\,\ref{fig:plot.density.figures.complete.mix.stackI} for example, we 
crossed a Feshbach resonance while increasing $V$, but no sharp density profile 
indicative of a collapse can be seen. In the next section, we shall explain the 
occurance of the negative energies to be the result of a liquefaction process of 
the Bose gas since when the energy becomes negative, energy is released from the 
system.

\subsubsection{Liquefaction of Bose gases (negative energies)}

\hs The bosons can condenste to a dense liquid at critical values of the 
scattering length and potential depth. This is signified by the occurence of 
negative energies as in Figs.\,\ref{fig:plot.Evsalpha}, 
\ref{fig:plot.Evs.alpha.1BEC.HCSW.N20.two.minima.special}, and 
\ref{fig:plotEvsalpha.R0d25.R0d54}. The liquid density is set by the range of 
the attractive well since there is a large energy incentive for a boson to lie 
in the attractive well of its neighbours but no further potential energy 
incentive to lie closer than that. There is a cost in kinetic energy increase if 
the bosons move closer together. Thus the density of the liquid saturates 
eventually to a value set by the range of the HCSW (and less so by the HC 
diameter).

\hs There is a large energy release on condensation of the Bose gas to the dense 
liquid state. In Fig.\ref{fig:plotEvsalpha.R0d25.R0d54} we show how the energy 
drops from $E/N\approx 1.6$ at $V_{12}=0$ to $E/N \approx -90$ at 
$V_{12}=-10.0$. Thus energy is released at a degree that is proportional to the 
well depth $V$. Upon condensation, there is substantial depletion of the 
condensate so that a large percentage of the atoms lie in states above the 
condensate. Thus the condensation process is characterized by large increase in 
the density of the system, a large drop in the energy of the system until there 
is a large release of condensation, and a substantial depletion of the condensate.

\subsection{Condensate depletion}

\hs In this section we explore possible reasons for the enhanced depletion of 
the condensates in the mixtures. We believe that due to mixing the reduction in 
free volume between the HS bosons available for condensate formation in each 
species is a common ground for enhanced depletion in all types of mixtures. The 
magnitude of reduction in free volume varies however with the types and 
strengths of interactions. As the free volume between the HS bosons decreases, 
the probability for relocating a boson at a certain site, say $\mathbf{r_1}$
and energy $\epsilon_0$ to another location at $\mathbf{r_1}^\prime$ and the 
same energy $\epsilon_0$ is reduced. This is because the chance of finding a 
site between the bosons large enough to accommodate a boson becomes lower.

\hs Going back to Fig.\ft\ref{fig:plot.n01andn02.for.intra.and.inter.HCSW.2BECs} 
then, the reason the mixtures with interspecies attraction (of 
Fig.\ft\ref{fig:plot.density.figures.complete.mix.stackI}) indicated by (inter.) 
show a larger depletion for component 1 (with HC $a_c$) than the mixtures with 
interspecies repulsion (of 
Fig.\ft\ref{fig:plot.densityprofiles.with.attrctve.intra.8FIGSSTACKB}) indicated 
by (intra.) is because the former are completely mixed as compared to partial 
mixing of the latter. In the case of complete mixing the available volume for 
condensate formation is severely reduced and smaller than the case of partial 
mixing. Further, since both components in 
Fig.\ft\ref{fig:plot.density.figures.complete.mix.stackI} are localized at the 
center of the trap, they contribute to their mutual condensate depletion where 
the density is highest, namely at the center of the trap. In the case of partial 
mixing the shell contributes to the depletion chiefly at the edges of the 
trap and does not influence the condensate at the center of the trap very much. 
There could also be other reasons that explain the enhancement in the depletion.

\hs The scenario is however different for component 2. In 
Fig.\ft\ref{fig:plotN10mixedHCandHCSWand1BECcomparisons} the depletion is larger 
for the 2BEC with interspecies repulsion (of 
Fig.\ft\ref{fig:plot.density.figures.HC.mix.stackI}) than the 2BEC with 
interspecies attraction (of 
Fig.\ft\ref{fig:plot.density.figures.complete.mix.stackI}). The reason may be 
because the boson sizes $a_c$ of the HC 2BEC are increased far above those of 
the HCSW 2BEC which remain fixed thus outweighing the role of the HCSW in the 
depletion (free-volume reduction).

\subsection{Ground state solutions}

\hs Trippenbach \ea\ \cite{Trippenbach:00} identified all possible classes of 
solutions for 2BECs and found that, in the case of isotropic harmonic trapping 
potentials, many spherically-symmetric phase-separated geometries are possible. 
In addition, symmetry breaking solutions do exist but within the TF 
approximation the ground state cannot be one with broken symmetry. Similarly, 
our mixtures are spherically symmetric in their ground states. 

\section{Conclusions}

\hs In summary then, we have investigated the effect of intra- and interspecies 
interactions on the properties of ultracold 2BECs in tight harmonic traps using 
VMC. The repulsive inter- or intraspecies interactions were modelled by a HC 
contact potential, the radius of which being equivalent to the s-wave scattering 
length in the low-energy and long-wavelength approximation. The attractive 
interactions were modelled by a HC repulsive part plus a shallow attractive 
well, the HCSW. We did not describe the attractive interactions by the HCSW 
scattering length $a$, but rather the depth of the HCSW in order to avoid the 
large fluctuations in the value of $a$. We calculated the energies, density 
distributions and condensate density distributions. We further obtained the 
condensate fractions of the components from the OBDMs. A key point is that we 
chiefly focused on the role of interactions in enhancing the condensate depletion 
of each component in a mixture as compared to the case when each component is in 
a {\it separate trap of its own}. To the best of our knowledge, this has not been 
done in the previous literature on mixed Bose gases up to this date. We present 
novel physics associated with the effect of intra-and interspecies interactions 
on the condensate properties of mixed Bose gases. We find that:

\begin{itemize} 
\item[a)] the mixing of two Bose gases in a trap enhances the condensate 
depletion of each gas as compared to the case when either one is in a separate 
trap of its own. In both cases of attractive and repulsive interactions the 
reduction in the available volume for condensate formation due to mixing plays a 
key role in the enhancement of the depletion. In the case of attractive 
interspecies interaction the enhanced depletion may be further driven by the 
liquefaction of the Bose gases at the higher densities and the release of 
energy.

\item[b)] when the condensates are phase separated due to strong repulsive 
interspecies interactions, the core remains stable and is not ``squeezed" 
substantially by the shell. Complete mixing is still possible up to some 
repulsion threshold.

\item[c)] according to Refs. \cite{Trippenbach:00}, our mixtures are stable 
because they are spherically symmetric.

\item[d)] we anticipate that the healing length of a 1BEC changes upon mixing it 
with a second component into the system.

\item[e)] in the case of complete phase-separation, although the density of the 
shell is much smaller than the core, substantial depletion is still observed in 
the shell triggered by the presence of large bosons in the shell.

\item[f)] in the case of intercomponent attraction a 2BEC behaves similarly to a 
1BEC as the two components are completely mixed and allow the condensates of 
either component to migrate towards the edges of the trap at the higher 
densities. This is contrary to the somewhat opposite case of repulsive 
interspecies interactions where the condensate of the core is pushed back 
towards the center of the trap.

\item[g)] finally the HC potentials provide a strong stabilizing mechanism for 
the Bose gases with attractive interactions.

\end{itemize}

\begin{acknowledgements}
\hs This work was partially funded by the NSF. We thank Humam B. Ghassib and 
William J. Mullin for a critical reading of the manuscript.
\end{acknowledgements}

\begin{appendix} \label{Appendix}
\section{Mean-Field Model for the Estimation of the Mixture-Energies}

\hs We consider two mixed Bose gases of $N_1$ and $N_2$ particles, HC diameters 
$a_c$ and $b_c$, and bosonic masses $m_1$ and $m_2$, respectively, where 
initially the interspecies interactions are set to zero. That means the two Bose 
gases are initially independent of each other and both of them are concentric 
spheres at the trap center. We then construct a rough model that describes the 
energy of a boson-boson mixture by using the following assumptions. In our 
estimate for the energies, we derive our concepts from a paper by Ao and Chui 
\cite{Ao:98} who gave a simplified expression for the total energy of an 
inhomogeneous binary Bose gas,
\begin{equation}
E\,=\,\frac{1}{2}\left(G_{11}\frac{N_1^2}{V}\,+\,G_{22}\,\frac{N_2^2}{V}\,+\,
\sqrt{G_{11}G_{22}}\,\cdot\,\frac{N_1 N_2}{V}\right) \label{eq:EnAoandChui1998}
\end{equation}

where $G_{ij}\,=\,4 \pi \hbar^2 a_{ij}/m_{ij}$ are the interaction parameters, 
$a_{12}$ the inter- and $a_{ii}$ the intraspecies s-wave scattering lengths, 
$m_{ii}$ the mass of a boson in one component and $m_{12}$ the reduced mass and 
$V$ is the volume of the gas. This expression neglects the kinetic energy 
(quantum pressure) of each component. We modify this expression by replacing the 
first two terms on the right-hand-side by the Thomas-Fermi (TF) energy of each 
component. We use trap units in terms of component 1, i.e 
$a_{ho}=\sqrt{\hbar/m_1\omega_1}$ and $\hbar\omega_1$ for both systems as done 
before. The TF energy for each component is then
\begin{equation}
E_{TF,i}\,=\,\frac{5}{7}\,N_i\,\mu_i \label{eq:TFenergy1}
\end{equation}

where $i=1,2$, $\mu_1\,=\,\frac{1}{2}\,(15\,N_1\,a_c)^{2/5}$ is the chemical 
potential of component 1 and 
\begin{equation}
\mu_2\,=\,\frac{1}{2}\,(15\,N_2\,b_c)^{2/5}\,\left(\frac{m_2}{m_1}\right)^{1/5}
\label{eq:mu2}
\end{equation}

that of component 2. The TF radius of component 1 is 
$R_{TF,1}\,=\,(15\,N_1\,a_c)^{1/5}$ and that of component 2
\begin{equation}
R_{TF,2}\,=\,(15\,N_2\,b_c)^{1/5}\left(\frac{m_1}{m_2}\right)^{2/5}. 
\label{eq:RTF2}
\end{equation}

Imagine now switching the interspecies interactions on such that the Bose gas 
with the larger HC diameter is expelled towards the edges of the trap and forms 
a shell. The shell would then lie approximately at the TF radius of the core
away from the center of the trap. The volume of the cloud is then approximately 
$V\,=\,4\,\pi\,R_{TF,1}^3/3$ if $a_c<b_c$and $4\pi R_{TF,2}^3/3$ if $a_c>b_c$. 
Thus the size of the cloud in largely determined by the core in the case of 
moderate repulsive interactions. In order to calculate its energy, we therefore 
assume a superposition of its initial TF energy when both mutually 
noninteracting components are localized at the center of the trap and an 
approximate potential energy for the particles of the shell formed at the edges 
of the trap after switching on the interspecies interactions. As a result, 
component 2 gains additional potential energy beyond $E_{TF,2}$ when it becomes 
a shell. An estimate for the potential energy of the shell is

\begin{eqnarray}
V_{trap}\,&=&\,\frac{1}{2}\,N_1\,m_1\,\omega_{ho}^2\,R_{TF,2}^2/\hbar\omega_{ho}
\,=\,\nonumber\\&&\frac{1}{2}N_1\left(\frac{m_1}{m_2}\right)^{4/5}\left(15 N_2 
b_c\right)^{2/5} \label{eq:Vtrap1}
\end{eqnarray}

if component 1 forms a shell and
\begin{eqnarray}
V_{trap}\,&=&\,\frac{1}{2}\,N_2\,m_2\,\omega_{ho}^2\,R_{TF,1}^2/\hbar\omega_{ho}
\,=\,\nonumber\\&&\frac{1}{2}N_2\left(\frac{m_2}{m_1}\right)\left(15 N_1 
a_c\right)^{2/5} \label{eq:Vtrap2}
\end{eqnarray}

if component 2 forms a shell. That is we used
\begin{equation}
\frac{\sum_{i=1}^{N_1} m_1\,r_{1\,i}^2}{\sum_{i=1}^{N_1}\,m_1}\,=\,
\langle r_1^2 \rangle \,\approx\,R_{TF,2}^2 \label{eq:m.averagepos1}
\end{equation}

and\\
\vspace{-0.5cm}
\begin{equation}
\frac{\sum_{i=1}^{N_2} m_2\,r_{2\,i}^2}{\sum_{i=1}^{N_2}\,m_2}\,=\,
\langle r_2^2 \rangle \,\approx\,R_{TF,1}^2 \label{eq:m.averagepos2}
\end{equation}
 
in estimating the radius of gyration of the core. Note that 
$\sum_{i=1}^{N_k}\,m_k\,=\,N_k m_k$ ($k=1$ or 2) is the total mass of 
either component as all the particles in a component have the same mass. The 
interspecies interactions can be calculated according to 
\ref{eq:EnAoandChui1998}:
\begin{equation}
E_{int}\,=\,\sqrt{G_{11}G_{22}}\,\frac{N_1 N_2}{V} \label{eq:EintAoandChui}
\end{equation}

Gathering all the previous terms together, the total energy of the mixture is 
then
\begin{widetext}
\begin{eqnarray}
&&E^{MFA}\cdot (N_1+N_2)\,\approx\, \nonumber\\
&&\frac{5}{15}\left[N_1(15 N_1 a_c)^{2/5}\,+\,N_2(15 N_2 b_c)^{2/5} 
\left( m_2/m_1 \right)^{1/5}\right] +
3[a_c b_c (m_1/m_2)]^{1/2} \cdot N_1 N_2 \left\{{\begin{array}{r@{\quad:\quad}l} 
R^{-3}_{TF,1} & a_c<b_c \\R^{-3}_{TF,2} & a_c>b_c \end{array}} \right. \,+\, 
\nonumber \\&&\left\{{\begin{array}{r@{\quad:\quad}l} \frac{1}{2}N_1 \left( 
m_1/m_2 \right)^{4/5} (15 N_2 b_c)^{2/5} & a_c > b_c\\\frac{1}{2}N_2\left( 
m_2/m_1 \right) (15N_1a_c)^{2/5} & a_c < b_c  \end{array}} \right. 
\label{eq:E_MFA}
\end{eqnarray}
\end{widetext}

where as a reminder $\omega_1\,=\,\omega_2$.

\end{appendix}

\bibliography{./Mixtures}

\end{document}